\documentclass[acmsmall]{acmart}

\usepackage{balance}
\usepackage{graphicx}
\usepackage{algorithm}
\usepackage{algorithmic}
\usepackage{amsmath}
\usepackage{multirow}
\usepackage{xcolor}
\usepackage{color}

\usepackage{amsmath}
\usepackage{float}
\usepackage{indentfirst}
\usepackage{subfigure}
\usepackage{flushend}
\usepackage{bm}

\usepackage{amssymb}
\usepackage{flushend}
\usepackage{setspace}

\def\BibTeX{{\rm B\kern-.05em{\sc i\kern-.025em b}\kern-.08emT\kern-.1667em\lower.7ex\hbox{E}\kern-.125emX}}
   
\definecolor{revise}{RGB}{0,0,255}   
\definecolor{revise_major}{RGB}{40,200,40}   
 
\copyrightyear{2020}
\acmYear{2020}
\setcopyright{acmlicensed}
\acmConference[Woodstock '18]{Woodstock '18: ACM Symposium on Neural Gaze Detection}{June 03--05, 2018}{Woodstock, NY}
\acmBooktitle{Woodstock '18: ACM Symposium on Neural Gaze Detection, June 03--05, 2018, Woodstock, NY}
\acmPrice{15.00}
\acmDOI{XXXXXXXXXXXXXXXXX}
\acmISBN{XXX-X-XXXX-XXXX-X/XX/XX}

\begin{document}

%
\title{Cross-Technology Communication for the Internet of Things: A Survey}

\author{Yuan He}
\affiliation{%
  \institution{Tsinghua University}
  \streetaddress{30 Shuangqing Rd}
  \city{Haidian}
  \state{Beijing}
  \country{China}}
\email{heyuan@mail.tsinghua.edu.cn}

\author{Xiuzhen Guo}
\affiliation{%
  \institution{Tsinghua University}
  \streetaddress{30 Shuangqing Rd}
  \city{Haidian}
  \state{Beijing}
  \country{China}}
\authornote{Xiuzhen Guo is the corresponding author.}
\email{guoxiuzhen94@gmail.com}

\author{Xiaolong Zheng}
\affiliation{%
  \institution{Beijing University of Posts and Telecommunications}
  \streetaddress{10 Xitucheng Rd}
  \city{Haidian}
  \state{Beijing}
  \country{China}}
\email{zhengxiaolong@bupt.edu.cn}

\author{Zihao Yu}
\affiliation{%
  \institution{Tsinghua University}
  \streetaddress{30 Shuangqing Rd}
  \city{Haidian}
  \state{Beijing}
  \country{China}}
\email{zh-yu17@mails.tsinghua.edu.cn}

\author{Jia Zhang}
\affiliation{%
  \institution{Tsinghua University}
  \streetaddress{30 Shuangqing Rd}
  \city{Haidian}
  \state{Beijing}
  \country{China}}
\email{j-zhang19@mails.tsinghua.edu.cn}

\author{Haotian Jiang}
\affiliation{%
  \institution{Tsinghua University}
  \streetaddress{30 Shuangqing Rd}
  \city{Haidian}
  \state{Beijing}
  \country{China}}
\email{jht19@mails.tsinghua.edu.cn}

\author{Xin Na}
\affiliation{%
  \institution{Tsinghua University}
  \streetaddress{30 Shuangqing Rd}
  \city{Haidian}
  \state{Beijing}
  \country{China}}
\email{nx20@mails.tsinghua.edu.cn}

\author{Jiacheng Zhang}
\affiliation{%
  \institution{Tsinghua University}
  \streetaddress{30 Shuangqing Rd}
  \city{Haidian}
  \state{Beijing}
  \country{China}}
\email{jc-zhang17@mails.tsinghua.edu.cn}

%
\renewcommand{\shortauthors}{X. Guo, et al.}

%

\begin{abstract}
The ever-developing Internet of Things (IoT) brings the prosperity of wireless sensing and control applications. In many scenarios, different wireless technologies coexist in the shared frequency medium as well as the physical space. Such wireless coexistence may lead to serious cross-technology interference (CTI) problems, e.g. channel competition, signal collision, throughput degradation. Compared with traditional methods like interference avoidance, tolerance, and concurrency mechanism, direct and timely information exchange among heterogeneous devices is therefore a fundamental requirement to ensure the usability, inter-operability, and reliability of the IoT.
Under this circumstance, Cross-Technology Communication (CTC) technique thus becomes a hot topic in both academic and industrial fields, which aims at directly exchanging data among heterogeneous devices that follow different standards. 
This paper comprehensively summarizes the CTC techniques and 
reveals that the key challenge for CTC lies in the heterogeneity of IoT devices, including the incompatibility of technical standards and the asymmetry of connection capability.
Based on the above finding, we present a taxonomy of the existing CTC works (packet-level CTCs and physical-level CTCs) and compare the existing CTC techniques in terms of throughput, reliability, hardware modification, and concurrency.

\end{abstract}

%
%


%


%

%
\maketitle

\section{Introduction}
\label{sec:introduction}

Wireless communication is the key to connecting countless devices around the world. As IoT applications widely spread, wireless technologies will get proliferated everywhere~\cite{bib:WiFi_material, bib:MET, bib:tracking_platform, bib:PTrack, bib:liquid, bib:beacon_wild, bib:pavatar}. Note that IoT applications are born to be diverse, with respect to many different factors, e.g., the deployment and operational environment, system scale, communication range, energy budget, desired network bandwidth, etc~\cite{bib:WiFi_60G, bib:PDLens, bib:Carmap, bib:widesee, bib:lora_sensing, bib:RFID_gesture, bib:smogy, bib:palantir}. Wireless technologies for IoT are intrinsically diverse as well. \textit{One size doesn't fit all.} Every type of wireless technology fits a certain category of applications. A number of different wireless technologies therefore coexist in the IoT era~\cite{bib:sensor_smartphone, bib:bluetooth_smartphone}.

A big portion of wireless technologies operates at the industrial, scientific and medical (ISM) band, sharing limited spectrum resource. As more and more IoT devices are deployed, we expect to see the increasingly crowded wireless channel. 
It is a crucial and significant problem to make those coexisting wireless technologies efficiently coordinate or even cooperate over the shared spectrum. 

Conventional approaches to tackle the above problem are using schemes based on collision avoidance to separate different technologies in accessing the spectrum, such as TDMA~\cite{bib:TDMA_based, bib:distributed_TDMA} and CSMA~\cite{bib:Adaptive_duty, bib:MAC_wlan}.
Methods like TDMA require end devices to use a common communication standard, which doesn't work in the scenario involving different wireless technologies.
Methods like CSMA may effectively avoid collisions but have proved inefficient in spectrum resource utilization. 
Driven by the need for ubiquitous connectivity in the IoT era, we desire to share information more efficiently than how it is collected nowadays. 

CTC emerges as such a technology that enables direct communication among heterogeneous devices that follow different wireless standards~\cite{WEBee, bluefi, bib:wibeacon, wizig, bib:Scylla, bib:X_burst, bib:narrowband_decoding, bib:X_MIMO, bib:ToneSense, bib:WiRa, bib:YaoJunmei}. It works like a translator that builds a mutually compatible side channel between two wireless technologies. CTC not only creates a new way for inter-operation and data exchange among wireless devices but also enhances the ability to manage wireless networks. 

Many studies on CTC have been proposed to fulfill the function of translation and to support applications like channel coordination and cross-technology collaboration ~\cite{WEBee, bib:wibeacon, wizig, bib:Scylla, bib:X_burst, bib:narrowband_decoding, bib:X_MIMO, bib:ToneSense, bib:zhangjia_jrnl, bib:BiCord, bib:SmarTiSCH}. 
\textbf{Most of the existing works focus on the enabling technology of CTC, but a complete picture of the literature and the research space is missing so far.} Based on the survey of recent studies, our paper presents a taxonomy of the existing CTC works (packet-level CTCs and physical-level CTCs) and compares the existing CTC techniques in terms of throughput, reliability, hardware modification, and concurrency. The findings and summaries of our paper are potentially significant in the following directions: 1) to guide subsequent researchers to rethink the CTC techniques regarding the design methodology; 2) to further innovate the infrastructure of future IoT by introducing CTC; 3) to enable important IoT application by enabling ubiquitous network connectivity.


Although there are existing surveys (e.g.~\cite{anotherSurvey}) on CTC, our paper presents new results in the following aspects:
(1) Our paper is more inclusive and reflects the latest progress in the area of CTC. 
(2) We propose a different taxonomy of CTC. Our paper classifies the existing CTC works into two categories: packet-level CTCs and physical-level CTCs, from the perspective of how to resolve the heterogeneity of IoT devices. The survey in~\cite{anotherSurvey} classifies the existing CTC works straightforwardly: hardware-based CTCs and hardware-free CTCs, from the perspective of whether hardware modification is required.
(3) We conduct comprehensive comparison and analysis, presenting a full-scale understanding of the existing works on CTC. Specifically, we compare the performance of CTC techniques in terms of throughput, reliability, hardware modification, and concurrency. 
(4) The scope of study in our paper is broader than that in~\cite{anotherSurvey}. Besides introducing and discussing the specific CTC techniques, our paper also illuminates the upper-layer application scenarios and the cutting-edge directions, which may motivate more follow-up studies in this area.

The main contributions of this paper are summarized as follows: 

\noindent $\bullet$ This paper presents a complete picture of the literature in the area of CTC. 

\noindent $\bullet$ In this paper, we point out that the key challenge for CTC lies in the heterogeneity of IoT devices, including the incompatibility of technical standards and the asymmetry of connection capability.

\noindent $\bullet$ This paper envisions the upper-layer application scenarios and discusses the research space of CTC, which may inspire researchers to innovate the design of future IoT.

The remainder of this paper is structured as follows.
This survey first discusses the problem of wireless coexistence as the background of CTC in Section~\ref{sec:background}, and introduces the application of CTC in Section~\ref{sec:application}.
We then describe the packet-level CTC and physical-level CTC in Section~\ref{sec:packet} and Section~\ref{sec:physical}.
Section~\ref{sec:overhead} compares CTC techniques and Section~\ref{sec:upperlayer} presents the upper layer CTC.
We also discuss the future directions of CTC in Section~\ref{sec:future}, with a particular emphasis on how it will innovate the design of future wireless systems.
We conclude in Section~\ref{sec:conclusion}.

\section{Background}
\label{sec:background}
Studies on CTC are motivated by the need for effective cooperation and efficient data exchange between heterogeneous devices coexisting in the same scenario. This section discusses the background studies, including collision avoidance, collision tolerance, and indirect bridging approaches. Based on such discussion, we better understand why CTC is needed and what problems CTC exactly addresses.

\subsection{Collision Avoidance}
Some approaches to avoid collision among heterogeneous devices are similar to those for homogeneous devices. We can leverage the methods of CSMA, TDMA, and frequency hopping to avoid the collisions~\cite{bib:whitespace, bib:bluetooth, bib:afhimprove, bib:wlanmodel, bib:arch}. But in scenarios where heterogeneous devices coexist, because of the asymmetry of communication ability, computational ability, and other abilities, the data transmission of the weaker device may not be perceived by the stronger device. 
For example,  the weaker device with lower transmission power is in the communication range of the stronger device with higher transmission power, but not vice versa. 
The stronger device hardly knows the weaker device's existence and has a significant impact on its data transmission.
Therefore, in scenarios where heterogeneous devices coexist, the weaker ones need to avoid the collision actively.

\subsection{Collision Tolerance}

Researchers focus on error detection and recovery methods to achieve the collision tolerance~\cite{bib:cognitive, bib:BLU, bib:ZiMo, bib:capture_effect, bib:fireworks}. PPR \cite{bib:PPR} is a partial retransmission method that proposes retransmission of the parts with errors instead of the whole packet to save cost. In \cite{bib:ewsn2010, bib:buzzbuzz}, the authors find that the bit errors of a ZigBee packet are temporally correlated with 802.11 traffic. 
Coding~\cite{bib:xors, bib:analogcoding} is another method to tolerate the interference. Rateless coding \cite{bib:strider,bib:automac} is a recent coding technique that allows the sender to transmit linearly-independent coded packets without knowing the channel condition as a prior. The receiver can decode the packets after accumulating enough correctly-coded packets.

\subsection{Indirect Bridging}
Different wireless technologies can be bridged by building an indirect connection through a multi-radio gateway. With multiple wireless interfaces equipped, a gateway can translate data from different devices following different communication standards. There are many applications that use the gateway to perform the data processing \cite{gateway_problem, Vivek2015EnablingIS, 2015Smart}. For example, Shim et al. \cite{2017Design} designs a ZigBee-BLE gateway to control and manage home appliances. These IoT systems implement multi-radio gateways that introduce additional hardware, maintenance costs, deployment complexity, and traffic overhead.

\subsection{Direct Connection}

CTC opens a new direction which enables direct communication among heterogeneous wireless devices~\cite{bib:free_side, bib:reactive_jamming, bib:correction, bib:Boosting}. The ability to communicate with devices of different technologies avoids the unnecessary hardware cost and communication delay, compared to the indirect solutions based on a multi-radio gateway \cite{gateway_problem}. With CTC, it becomes easier for heterogeneous wireless devices to coordinate even in a shared channel \cite{Mac}. CTC is also an enabling technology for emerging IoT applications (e.g., industrial surveillance and smart home), where seamless data collection and interoperation are desired \cite{bib:hanShop, emf, bib:zhenjiangPtrack}.
Therefore, CTC brings benefits to interference management, in-situ data exchange, and inter-operation among wireless devices.

\section{Application}
\label{sec:application}
With the rapid development of CTC links among heterogeneous devices, researchers have proposed many application scenarios to bring this technique into reality. By enabling direct communication among multiple coexisting technologies, CTC benefits existing wireless networking and our daily life in many aspects.

\begin{figure}[!tb]
\centering
\subfigure{\label{fig:gbee_a}}\addtocounter{subfigure}{-2}
\subfigure[ZigBee and WiFi channel layout]{\subfigure
{\includegraphics[width=2.0in]{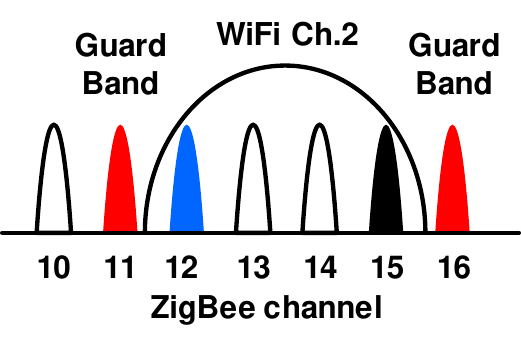}}}
\subfigure{\label{fig:gbee_b}}\addtocounter{subfigure}{-2}
\subfigure[G-Bee procedure]{\subfigure
{\includegraphics[width=3.2in]{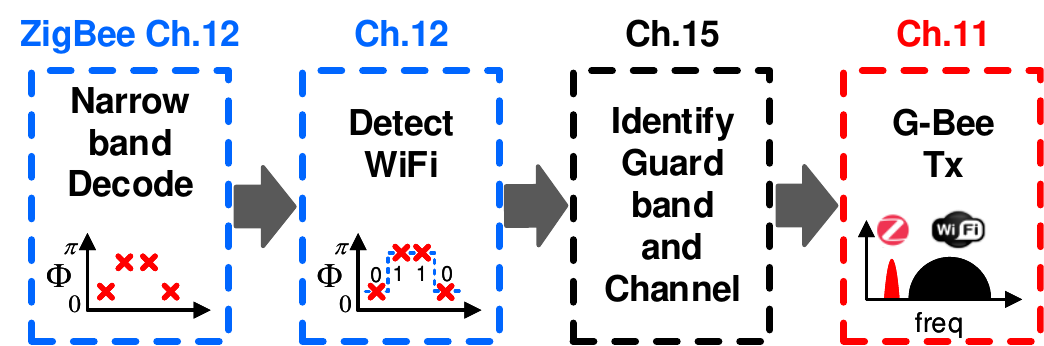}}}
\caption{Channel layout and operation procedure of G-Bee}
\vspace{-0.1in}
\label{fig:gbee}
\end{figure}

\subsection{Avoiding Cross-technology Interference}

\subsubsection{Passive avoidance}
With the ability to decode data packets of another technology, reliable communication can be established by learning free bands in the channel and therefore avoids CTI. Based on this idea, \textbf{G-Bee} \cite{GBee} leverages the \emph{guard band} of ongoing WiFi traffic to transmit ZigBee packets reliably. Guard bands separate WiFi channels by 3 or 5 MHz to avoid inter-channel interference. It is enough to accommodate a ZigBee transmission which requires 2 MHz bandwidth. As shown in Fig. \ref{fig:gbee}, ZigBee channel 11 and 16 can be safeguarded when WiFi transmission is in channel 2. Evaluation results show that G-Bee enhances packet reception rate (PRR) from 15\% to over 95\% under significant WiFi interference.

\subsubsection{Active avoidance}
Different from listening to WiFi transmissions passively, the active avoidance approach will back off WiFi transmissions and reserve certain transmission windows for ZigBee. Based on this idea, researchers in \textbf{ECC}~\cite{Mac} found that it is possible to aggregate white spaces in WiFi transmissions as the channel is idle for 230 ms in every second. They use a CTS message to clear the channel and notify ZigBee nodes to transmit by adopting CTC from WiFi to ZigBee. As a result, ECC achieves 1.8x ZigBee packet reception ratio under heavy interference of WiFi.

\subsection{Improving IoT Network Efficiency}
\begin{figure}[!tb]
\centering
\includegraphics[width=5.0in]{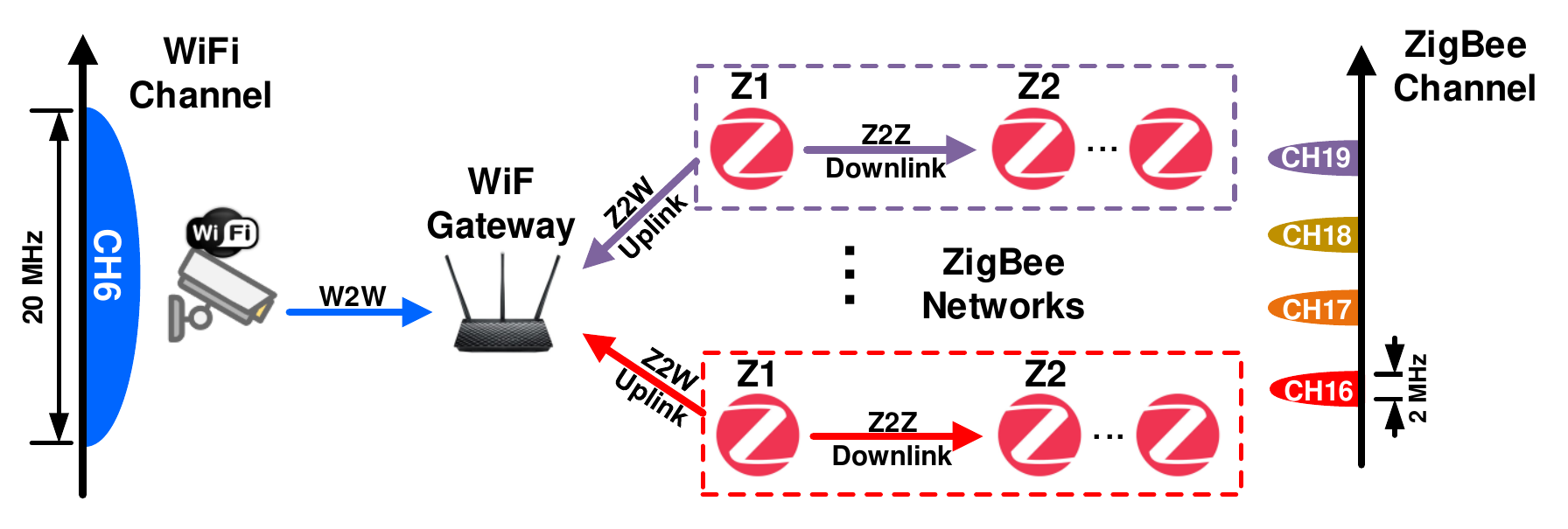}
\caption{Amphista network architecture. By using the CTC link from ZigBee to WiFi (Z2W), control messages and software updates are transmitted inside ZigBee networks. Overall, Amphista achieves concurrent communications of 1) WiFi high quality (HQ) video uploading, 2) four groups of ZigBee sensor data uploading, 3) four groups of ZigBee intra-communication.}
\label{fig:Amphista}
\end{figure}

In order to improve the network efficiency when IoT devices coexist with WiFi stations, \textbf{ECT}~\cite{ect} intends to reduce packet collisions by leveraging concurrent transmission and setting priorities for different nodes. As another example of network efficiency improvement with the help of CTC, \textbf{Amphista}~\cite{Amphista} presents a novel and effective approach to utilize the 2.4 GHz spectrum. As shown in Fig.~\ref{fig:Amphista}, by leveraging a single ZigBee stream, Amphista enables the ZigBee device to send different information to WiFi and ZigBee devices. The ZigBee to WiFi (Z2W) link embeds sensor data into ZigBee's transmission power. The same stream is used to disseminate software updates and control messages between ZigBee devices, i.e. Z2Z links. Amphista achieves highly concurrent transmissions and data forwarding of a WiFi video uploading, four groups of ZigBee to WiFi uploading, and four groups of ZigBee intra-communication. The experimental results show that Amphista improves network throughput by up to 400x and reduces transmission latency.

\subsection{Working with Backscatter Networking}

\begin{figure}[!tb]
\centering
\subfigure{\label{fig:passiveZigBee_nx_a}}\addtocounter{subfigure}{-2}
\subfigure[Sensor data transmission]{\subfigure
{\includegraphics[width=2.5in]{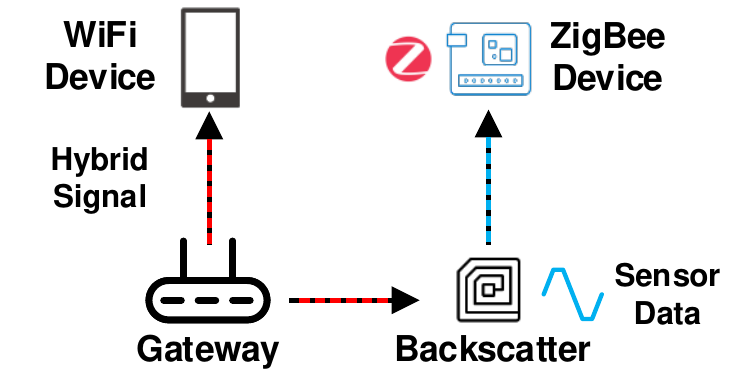}}}
\subfigure{\label{fig:passiveZigBee_nx_b}}\addtocounter{subfigure}{-2}
\subfigure[Backscatter act as relay]{\subfigure
{\includegraphics[width=2.5in]{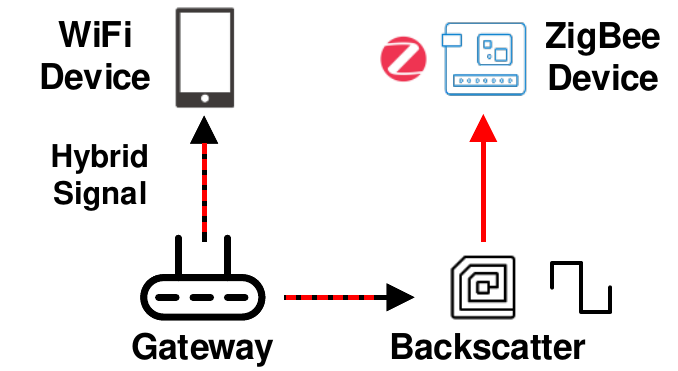}}}
\caption{PassiveZigBee overview. (a) Tag backscatters the signal to transmit sensor data. (b) Tag relays messages to ZigBee device.}
\label{fig:passiveZigBee_nx}
\end{figure}

As the rapid growth of IoT devices and sensors, ultra-low power and energy harvesting sensors are gaining more attention in areas such as health monitoring. A simple example would be energy-harvesting sensors, which are implanted in patients and collect data like electrocardiogram, that need to transmit the data to wearable ZigBee devices. The wearable devices also require localization and control data from a cloud server. Considering the energy consumption of traditional ZigBee and the repetitive packets overhead of bridge gateway, existing methods are unable to provide efficient transmission. In \cite{passivezigbee}, researchers propose to modify the WiFi payload of the gateway so that it can transmit a hybrid WiFi and ZigBee signal. As shown in Fig.~\ref{fig:passiveZigBee_nx}, by leveraging productive WiFi packets, a tag backscatters the hybrid signal to transmit its sensor data as well as relay messages to the ZigBee node. The tag changes the frequency of the hybrid signal with power consumption of only 25 $\mu$W. It demonstrates the feasibility and possible improvement of leveraging CTC in backscatter networking.

\subsection{Clock Synchronization}

Clock synchronization is a critical function in IoT networks, especially in event-driven scenarios such as environmental surveillance. \textbf{Crocs}~\cite{crocs} is the first work to synchronize clocks of WiFi and ZigBee devices. It consists of two phases. First, a WiFi device transmits beacon messages to form a detectable pattern and let the ZigBee device record timestamps of every received pattern. After agreeing on a unique global time point, timestamps recorded by the WiFi device are sent to ZigBee in the CTC link. Then, clock calibration and adjustment are conducted on ZigBee devices. Crocs provides a robust and accurate synchronization between WiFi and ZigBee with an error of less than 1 millisecond.

\subsection{Cross Technology Attack}
Although researchers have proposed many CTC links and applications in IoT networks, CTC transmission is vulnerable to network attacks such as jamming and sniffing~\cite{bib:Wi-attack}. Different from attacks on the existing network, reactively jamming packet-level CTC link requires prior knowledge of specific modulation scheme (e.g., packet pattern, power level). To this end, \textbf{JamCloak} \cite{jamcloak} proposes a taxonomy on different CTC protocols and trains a detection model to classify CTC traffic. Then, it conducts a reactive jamming attack on the CTC traffic. The results show that JamCloak can reduce the packet delivery ratio by 80.8\% of existing CTC links in practice. In addition, a countermeasure against such attacks is proposed and shown to be capable of resolving the jamming attack.

\section{Packet level CTC}
\label{sec:packet}

Early works on CTC utilize the packet transmission as the carrier to convey messages to the receiver of another technology. These works are named as packet level CTC. The sender usually transmits legitimate packets in an elaborate manner to encode information, which is decoded by the receiver based on the channel analysis. 
The existing works on packet level CTC are shown in Table~\ref{table: existing CTCs}. In the column of throughput, the values in the parentheses refer to the aggregated throughput when multiple senders or receivers are involved.

\begin{figure}[!tb]
\centering
\includegraphics[width=3.3in]{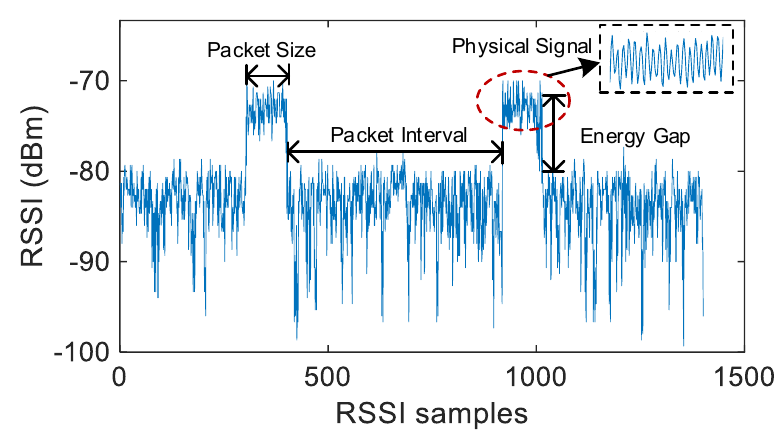}
\caption{RSSI as the side channel for packet-level CTC} 
\label{fig:plc_rssi}
\end{figure}

There are two typical ways that the receiver captures the characteristics of the sender's packet transmission and decodes the intended information. 
First, the receiver may utilize the received signal strength indication (RSSI) to recognize and analyze the packets from a heterogeneous device. As a foundational function for MAC techniques including CSMA, RSSI sampling is common among varieties of wireless standards (e.g., Bluetooth, ZigBee). Based on the fact that common technologies like WiFi, ZigBee and Bluetooth coexist on the 2.4GHz ISM band, one device is able to sense the energy of packets from another device in the channel if their frequency bands overlap. Therefore, RSSI can serve as a bridge between heterogeneous devices.
The second method utilizes channel  state  information  (CSI) as a side  channel to convey CTC symbols. This method is specific to the scenario where WiFi is the receiver. WiFi is able to sense the transmission of heterogeneous wireless devices like Bluetooth or ZigBee devices by analyzing the CSI values of the subcarrier of corresponding frequency. So these devices may convey information to WiFi by influencing the CSI values of ongoing WiFi transmissions. This method develops quickly in recent years as WiFi devices are widely deployed in various environments. 

\begin{table*}[]
\caption{Packet level CTC and physical level CTC}
\label{table: existing CTCs}
\resizebox{.95\columnwidth}{!}{
\begin{tabular}{|l|l|l|l|l|l|l|}
\hline
Category                                                                                         & Method                        & Link                                                     & Parallel CTCs & Throughput     & Reliability & Complexity \\ \hline
\multirow{3}{*}{\begin{tabular}[c]{@{}l@{}}RSSI-based CTC \\ using packet energy\end{tabular}}   & Basic \cite{dcoss}            & WiFi $\rightarrow$ ZigBee                                & No   & 16bps          & High   & Low      \\ \cline{2-7} 
                                                                                                 & WiZig \cite{wizig, bib:wizig_jrnl}            & WiFi $\rightarrow$ ZigBee                                & No   & 153.85bps      & High  & Low     \\ \cline{2-7} 
                                                                                                 & StripComm \cite{StripComm}    & WiFi $\rightarrow$ ZigBee                                & No  & 1.1kbps        & High  & Low       \\ \hline
\multirow{2}{*}{\begin{tabular}[c]{@{}l@{}}RSSI-based CTC \\ using packet size\end{tabular}}     & Esense \cite{Esense}          & WiFi $\rightarrow$ ZigBee                                & No   & N/A            & High  & Low     \\ \cline{2-7} 
                                                                                                 & HoWiES \cite{howies}          & WiFi $\rightarrow$ ZigBee                                & No   & N/A            & High   & Low      \\ \hline
\multirow{5}{*}{\begin{tabular}[c]{@{}l@{}}RSSI-based CTC \\ using packet schedule\end{tabular}} & FreeBee \cite{freebee}        & \begin{tabular}[c]{@{}l@{}}WiFi $\leftrightarrow$ ZigBee\\ Bluetooth $\rightarrow$ WiFi/ZigBee\end{tabular} & Yes       & 31.5(560)bps   & High   & Low      \\ \cline{2-7} 
                                                                                                 & DCTC \cite{dctc}              & WiFi $\leftrightarrow$ ZigBee                            & Yes       & 47-160(760)bps & High  & Low     \\ \cline{2-7} 
                                                                                                 & Gap Sense \cite{gapsense}     & WiFi $\rightarrow$ ZigBee                                & No   & N/A            & High  & Low       \\ \cline{2-7} 
                                                                                                 & C-Morse \cite{cmorse}         & WiFi $\rightarrow$ ZigBee                                & Yes       & 12-137(936)bps & High  & Low     \\ \cline{2-7} 
                                                                                                 & EMF \cite{emf}                & WiFi $\leftrightarrow$ ZigBee                            & Yes       & 203(356)bps    & High  & Low     \\ \hline
\begin{tabular}[c]{@{}l@{}}RSSI-based CTC \\ using packet content\end{tabular}                   & LoraBee \cite{shi2019lorabee} & LoRa $\rightarrow$ ZigBee                                & No   & 281.61bps      & High   & Low      \\ \hline
\multirow{6}{*}{CSI-based CTC}                                                                   & {$B^2W^2$} \cite{B2W2} DAFSK  & Bluetooth $\rightarrow$ WiFi                             & Yes       & 1.5kbps        & High   & Low    \\ \cline{2-7} 
                                                                                                 & ZigFi \cite{ZigFi, bib:zigfi_jrnl}            & ZigBee $\rightarrow$ WiFi                                & No   & 215.9bps       & High    & Low      \\ \cline{2-7} 
                                                                                                 & AdaComm \cite{AdaComm}        & ZigBee $\rightarrow$ WiFi                                & No  & 229bps         & High   & Low      \\ \cline{2-7} 
                                                                                                 & Amphista \cite{Amphista}      & ZigBee $\rightarrow$ WiFi                                & Yes       & 2500bps        & High   & Low    \\ \cline{2-7} 
                                                                                                 & cChirp \cite{cChirp, bib:cChirp_jrnl}          & ZigBee $\rightarrow$ WiFi                                & No  & 90.12bps       & High   & Low      \\ \cline{2-7} 
                                                                                                 & DopplerFi \cite{DopplerFi}    & Bluetooth $\leftrightarrow$ WiFi                         & No  & 1.59Kbps       & High   & Low      \\ \hline
\multirow{7}{*}{\begin{tabular}[c]{@{}c@{}}Receiver   \\ Transparent CTCs\end{tabular}}          & WEBee \cite{WEBee}            & WiFi $\rightarrow$ ZigBee                                & Yes       & 63 Kbps        & Low   & High        \\ \cline{2-7} 
                                                                                                 & PMC \cite{PMC}                & WiFi $\rightarrow$ ZigBee                                & Yes       & 121.02 Kbps    & Low   & High       \\ \cline{2-7} 
                                                                                                                                                                                                  & BlueFi \cite{bluefi}          & WiFi $\rightarrow$ Bluetooth                             & Yes       & 122.5  Kbps    & Low  & High        \\ \cline{2-7} 

                                                                                                 & LTE2B \cite{lte2b}            & LTE $\rightarrow$ ZigBee                                 & No   & /              & Low     & High     \\ \cline{2-7} 
                                                                                                 & Passive-ZigBee \cite{passivezigbee} & WiFi $\rightarrow$ ZigBee                          & No   & 230 Kbps       & Low   & High       \\ \cline{2-7} 
                                                                                                 & WIDE \cite{wide, bib:wide_jrnl}              & WiFi $\rightarrow$ ZigBee                                & Yes       & 247.2 Kbps     & Low   & High       \\ \cline{2-7} 
                                                                                                 & BlueBee \cite{bluebee}        & BLE $\rightarrow$ ZigBee                                 & No  & 225 Kbps       & Low   & High       \\ \hline
\multirow{3}{*}{\begin{tabular}[c]{@{}c@{}}Transmitter   \\ Transparent CTCs\end{tabular}}       & XBee \cite{xbee}              & ZigBee $\rightarrow$ BLE                                 & No  & 217 Kbps       & Low    & High      \\ \cline{2-7} 
                                                                                                 & LEGO-Fi \cite{legofi, bib:legofi_jrnl}         & ZigBee $\rightarrow$ WiFi                                & No   & 213.6 Kbps     & Low   & High       \\ \cline{2-7} 
                                                                                                 & XFi \cite{xfi}                & ZigBee $\rightarrow$ WiFi                                & Yes       & 285.7 Kbps     & Low    & High      \\ \hline
\multirow{6}{*}{\begin{tabular}[c]{@{}c@{}}None \\ Transparent CTCs\end{tabular}}                & TwinBee \cite{TwinBee}        & WiFi $\rightarrow$ ZigBee                                & Yes      & /              & Low   & High       \\ \cline{2-7} 
                                                                                                 & LongBee \cite{LoneBee}        & WiFi $\rightarrow$ ZigBee                                & Yes       & /              & Low    & High      \\ \cline{2-7} 
                                                                                                 & SymBee \cite{symbee}          & ZigBee $\rightarrow$ WiFi                                & No  & 31.25 Kbps     & Low     & High     \\ \cline{2-7} 
                                                                                                 & Chiron \cite{chiron}          & WiFi $\leftrightarrow$ ZigBee                            & Yes       & 223.97 Kbps    & Low    & High      \\ \cline{2-7} 
                                                                                                 & PIC \cite{pic}                & WiFi $\leftrightarrow$ BLE                               & Yes       & 121.02 Kbps    & Low    & High      \\ \cline{2-7} 
                                                                                                 & Symphony \cite{symphony}      & ZigBee, BLE $\rightarrow$ LoRa                           & Yes      & 3 Kbps         & Low    & High      \\ \hline
\end{tabular}}   

\end{table*}

\subsection{RSSI-based Packet Level CTC}
In RSSI-based packet level CTC, the sender and receiver communicate by producing and sensing the RSSI sequence of a certain pattern in the channel. Fig.\ref{fig:plc_rssi} shows how to establish a side channel based on RSSI between the sender and receiver which utilize heterogeneous technologies but coexist in an overlapping frequency band. Typically, when the sender sends packets in the channel, the receiver senses the existence of them by observing the fluctuation of RSSI values. In Fig.\ref{fig:plc_rssi}, the receiver is able to recognize two packets from the sender by analyzing these RSSI samples. In a subsequent analysis of RSSI samples, the receiver can calculate the packet size and packet interval of the sender. These parameters can be manipulated by the sender to convey information. Moreover, the energy gap between RSSI values related to packets and the noise floor is an optional side channel. 

In this section, we introduce the representative CTC works where the sender encodes messages in RSSI by manipulating packet energy, packet size, packet schedule or packet content.

\subsubsection{Packet energy}
Although ZigBee cannot decode WiFi packets directly due to incompatible standards, ZigBee can detect the presence of WiFi packets by background energy sensing. This sheds light on the CTC based on packet energy. In the Ref. \cite{dcoss}, the author proposes a CTC method for WiFi to communicate with ZigBee using packet energy as the information carrier. Since that the channel overlapping is an essential condition for ZigBee to perceive WiFi signals, WiFi channel $N$ and ZigBee channel $N+11$ are used to accomplish the CTC. As a basic RSSI-based CTC using packet energy, this work achieves a data rate of 2 bytes per second with less than $10\%$ bit error rate.

\begin{figure}[!tb]
\centering
\includegraphics[width=3.5in]{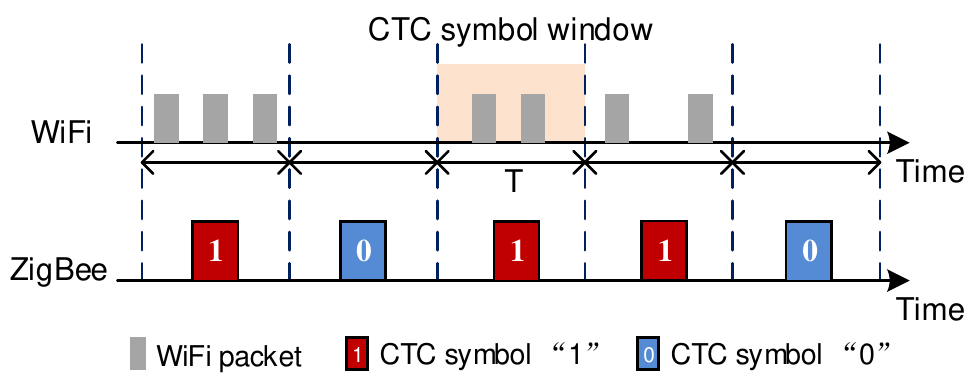}
\caption{The modulation and demodulation of the basic packet energy based CTC}
\label{fig:plc_dcoss}
\end{figure}

The modulation and demodulation scheme is shown in Fig. \ref{fig:plc_dcoss}. The WiFi sender modulates the energy in the channel by controlling the presence and the absence of high rate UDP packets. The presence and absence of high rate UDP packets represent the CTC symbol ``1'' and ``0'', respectively.
During the transmission of WiFi packets, the ZigBee receiver conducts energy sensing and collects the RSSI sequence. Then it uses the minimum RSSI fraction or average RSSI as an indicator to distinguish the presence or absence of the WiFi UDP packets. For example, the ZigBee receiver calculates the average RSSI within the decoding window. If this ratio is larger than the predefined threshold, the CTC symbol is decoded as ``1''. Otherwise, the CTC symbol is decoded as ``0''.

\begin{figure}[!tb]
\centering
\includegraphics[width=3.5in]{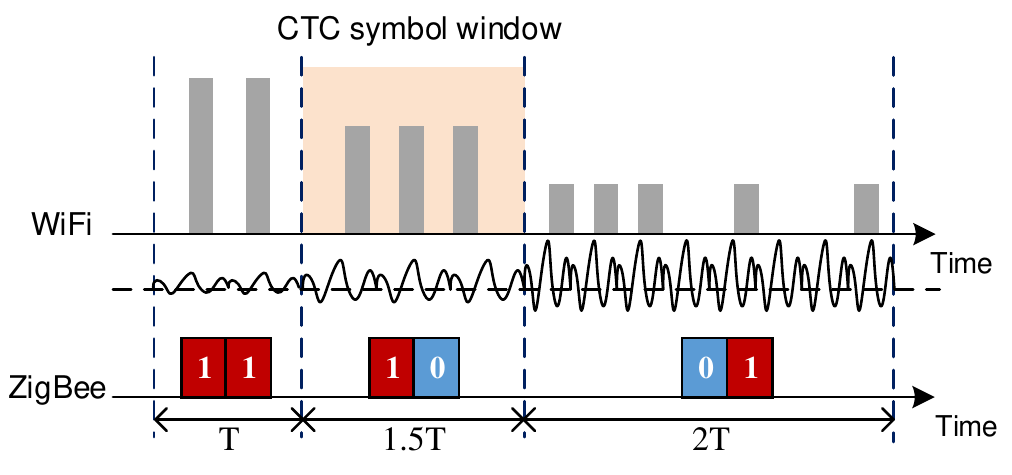}
\caption{WiZig with multiple energy levels and the adjustment of CTC symbol window length}
\label{fig:plc_wizig_window}
\end{figure}

It is inefficient to use only the presence and absence of WiFi packets to encode the CTC symbol, as there are only two energy levels (one binary CTC symbol) within a CTC window. To improve CTC throughput, \textbf{WiZig} \cite{wizig, bib:wizig_jrnl} proposes to increase the number of energy levels to encode multiple CTC symbols within a CTC window. Generally, WiZig offers a throughput of $153.85$ bps with less than $1\%$ symbol error rate in a real environment.
To explain the modulation of WiZig, we take the scheme of four energy levels as an example. As shown in Fig. \ref{fig:plc_wizig_window}, the WiZig sender transmits WiFi packets with three different powers to provide three energy levels, which can be encoded as ``01'', ``10'', and ``11''. The absence of WiFi packets is encoded as ``00''. The WiZig receiver samples the RSSI sequence on the overlapping channel and detects four different energy levels. In this way, the WiZig receiver can decode two CTC symbols within a CTC decoding window. If the number of energy levels is $M$, the CTC symbols with a CTC window are equal to $log_2M$.


\subsubsection{Packet size}
\textbf{Esense} \cite{Esense} is an early work that uses packet size as the side channel to realize uni-directional CTC from WiFi to ZigBee. To encode CTC symbols, Esense builds an "alphabet set" by exploring a set of packet sizes that are distinguished from the normal packet size of WiFi communication. Each packet size in the ``alphabet set'' can represent a different piece of information conveyed from WiFi to ZigBee. \textbf{HoWiES} \cite{howies} is another RSSI-based packet level CTC that manipulates packet size. HoWiES is inspired by Esense and expands the message capacity of CTC. 
In the Ref. \cite{WiFi_interrupt}, the author proposes a bi-directional CTC between WiFi and ZigBee. 

\subsubsection{Packet schedule}
\begin{figure}[!tb]
\centering
\includegraphics[width=3.5in]{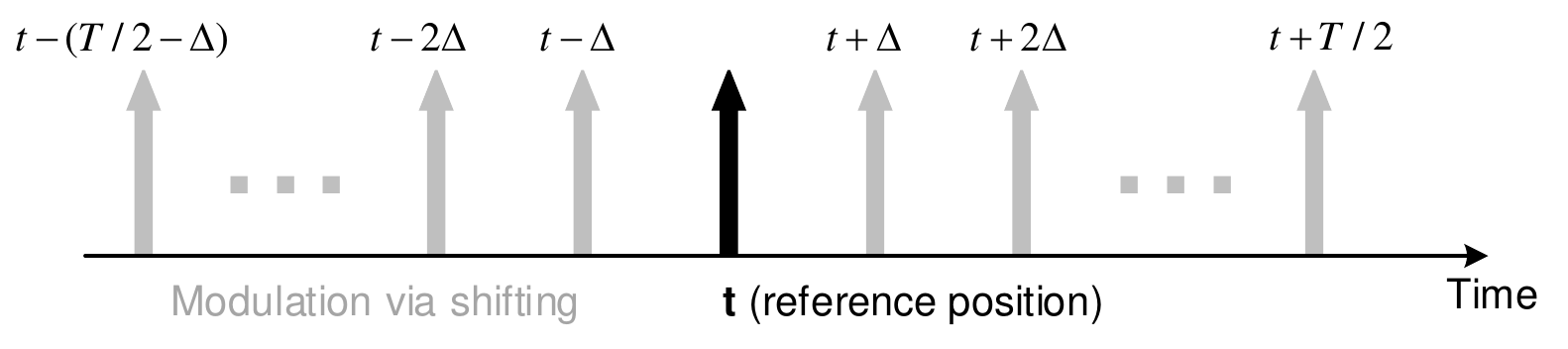}
\caption{The modulation of basic FreeBee}
\label{fig:plc_freebee}
\end{figure}

In the Ref. \cite{freebee}, the author proposes \textbf{FreeBee}, a CTC framework among three popular wireless technologies: WiFi, ZigBee, and Bluetooth. The key idea of FreeBee is to modulate CTC symbol messages by shifting the transmission timings of periodic beacons. FreeBee utilizes mandatory beacons widely adopted among wireless technologies and achieves a generic and free side channel design without incurring extra traffic.  We use the communication from WiFi to ZigBee to illustrate the generic design of FreeBee.
The modulation process of basic FreeBee is shown in Fig. \ref{fig:plc_freebee}. We assume that the reference position of the unmodulated beacon is at $t$ and the interval between two unmodulated beacons is $T$. The FreeBee sender shifts the beacon from its reference position in the range of $(\frac{-T}{2},\frac{T}{2}]$ to embed the CTC symbols. The capacity of FreeBee depends on the $T$ and the granularity of shift $\bigtriangleup$. Following the 802.11 standards, $\bigtriangleup$ is set as 1.024ms and the interval of the unmodulated beacons is 102.4ms. That means the beacon can be modulated at 100 different time instances and the beacon timing shift can represent 6 bits ($\lfloor log_2100 \rfloor$). The modulated beacons are transmitted several times repeatedly (i.e., beacons are transmitted at $t+T-\bigtriangleup$, $t+2T-\bigtriangleup$, and so on). The required number of beacon repetitions per CTC symbol is decided by the channel noise.

Different from FreeBee, \textbf{DCTC} \cite{dctc} achieves CTC with existing data packets, which significantly enhances the CTC throughput to 760bps. Besides modulating the timings of beacon and data packets, \textbf{Gap Sense} \cite{gapsense} constructs customized preambles and leverages the quiet period between the customized preambles to convey CTC symbols. The preamble is designed elaborately in a way to maximize the signal-to-noise ratio (SNR). 
The gap between two consecutive preambles can be determined by the receiver according to the number of samples that the receiver recorded between the RSSI pulses. Then the gap length can be mapped to various CTC symbols, depending on the objective of higher-layer protocols.
\textbf{C-Morse} \cite{cmorse} borrows the idea from International Morse Code, and uses the combination of the short WiFi packet (dot) and the long WiFi packet (dash) to construct recognizable energy patterns at the ZigBee receiver.
\textbf{EMF} \cite{emf} embeds CTC message by constructing different traffic occupancy ratios. The traffic occupancy ratio denotes the ratio of total packets duration to the total time duration. The change of packet occupancy ratio can be achieved by shifting the packets or flipping the packet order to form a unique pattern.

\subsubsection{Packet content}
\textbf{LoraBee}\cite{shi2019lorabee} explores how to enable CTC from LoRa to ZigBee devices. 
LoraBee observes that ZigBee can sense and capture the characteristics of LoRa communication by RSSI measurement.
When the LoRa and ZigBee channels overlap partially, ZigBee is able to detect the LoRa payload based on its RSSI signature. Therefore, the LoRa payload and the corresponding RSSI sequence measured by ZigBee can be used as a side channel to convey messages. Specifically, LoRaBee uses the sudden RSSI value drops as the feature to identify the RSSI signature of LoRa. This feature is resilient to noise and interference and can be captured by the receiving ZigBee node. As for the encoding scheme, the LoRa device puts the feature in the packet payload according to the conveyed CTC symbol, while the ZigBee device decodes the symbol when it detects the match between the measured RSSI feature and the feature set stored beforehand.
The experimental results show that LoRaBee provides reliable CTC communication from LoRa to ZigBee with a throughput of 281.61bps in the Sub-1 GHz band.

\subsection{CSI-based CTC}
CSI is normally used by WiFi to measure the channel variation during the WiFi packet transmission from the WiFi sender to the WiFi receiver. The CSI values include the phase and magnitude attenuation caused by channel changes at the subcarrier level and the CSI values can be obtained from the off-the-shelf WiFi devices(e.g., Intel 5300 card). Considering a WiFi device transmitting packets in the channel, the heterogeneous device can overlap its transmission at both the frequency band and the transmission time. That way, the device can influence the CSI values of the WiFi subcarriers of the same frequency and further, convey information by changing the CSI values. 

\begin{figure}[!tb]
\centering
\includegraphics[width=3.2in]{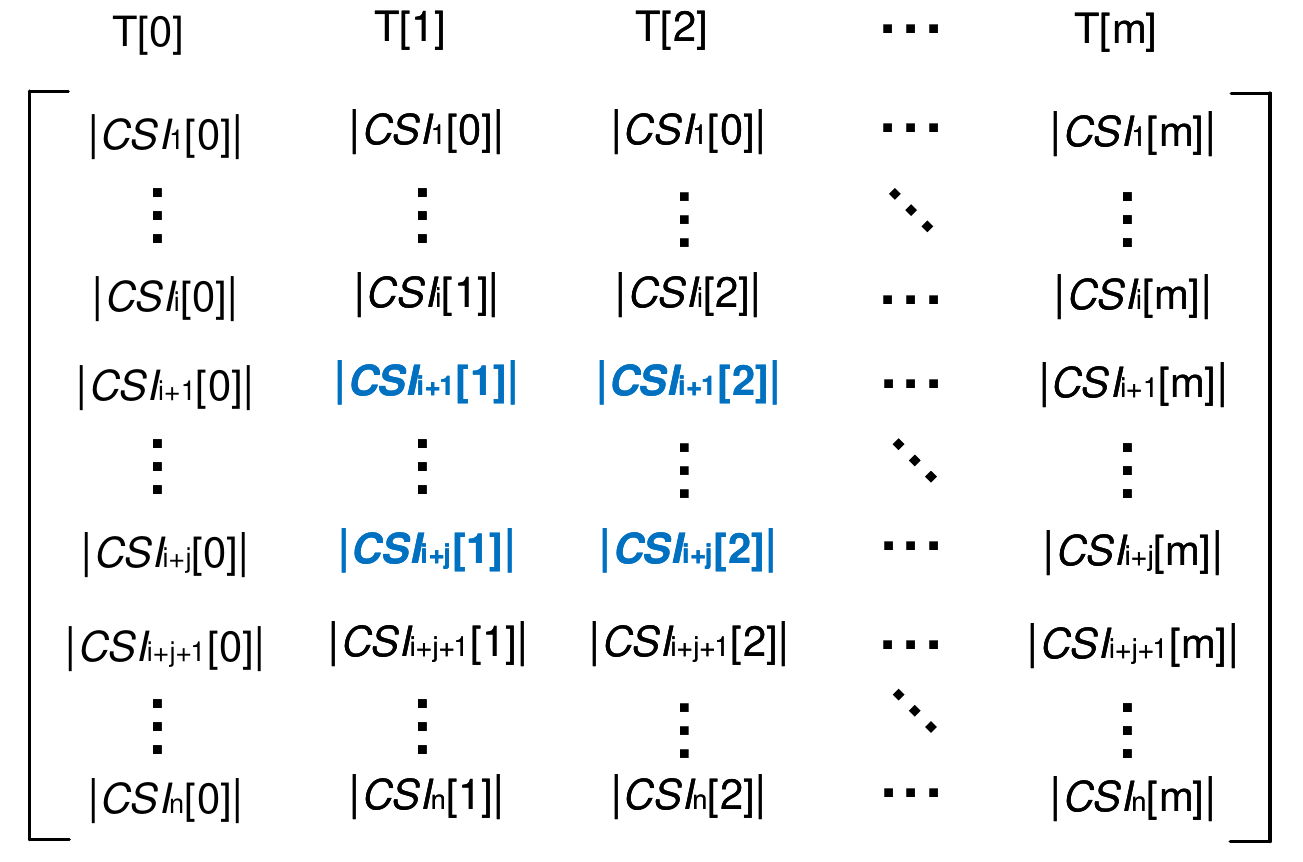}
\caption{CSI matrix from subcarrier 1 to $N$ after received WiFi packets at different time}
\label{fig:plc_csi}
\end{figure}

Specifically, the bandwidth of a WiFi subcarrier (312.5KHz) is several times narrower than one BLE or ZigBee channel (2MHz). In addition, a BLE or ZigBee packet is long enough to hit several WiFi packets in a row. Therefore, BLE or ZigBee devices can convey information by affecting the CSI values of WiFi subcarriers. At the WiFi side, considering a WiFi receiver that receives $m$ continuous packets at different time and the WiFi band is divided into $N$ subcarriers. The Fig. \ref{fig:plc_csi} shows the CSI matrix from time $T[0]$ to time $T[m]$ and from subcarrier $1$ to $N$. The blue-colored CSI values are affected by the transmission of heterogeneous devices. The researchers observe that, without the effect of other packets, the neighboring subcarriers (e.g., subcarrier i and i+1) should have very similar CSI readings due to the similar wavelength and multipath effect. So the WiFi device is able to extract the information embedded to the blue colored CSI values by calculating the following equation:
\begin{equation}
\label{eq:b2w2}
Y_{i+1}[k]=|CSI_{i+1}[k]|-|CSI_{i}[k]|
\end{equation}
where $Y_{i+1}[k]$ can be used to decode the information from a heteregeneous sender. 

Based on the analysis above, the CSI-based CTC achieves the message transfer to WiFi receiver from a heterogeneous sender. Recent works on CSI-based CTC also explore how to improve the reliability in dynamic conditions, extend the transmission range, avoid modifications to MAC-related configurations, or apply it in edge computing scenarios involving gateways.

\bm{{$B^2W^2$}} \cite{B2W2} enables CTC from BLE to WiFi while concurrently supports the original BLE to BLE and WiFi to WiFi communications. The basic idea of $B^2W^2$ is to leverage the CSI variation to embed the CTC symbol from BLE into its overlapped WiFi subcarriers. 

\begin{figure}[!tb]
\centering
\includegraphics[width=3.5in]{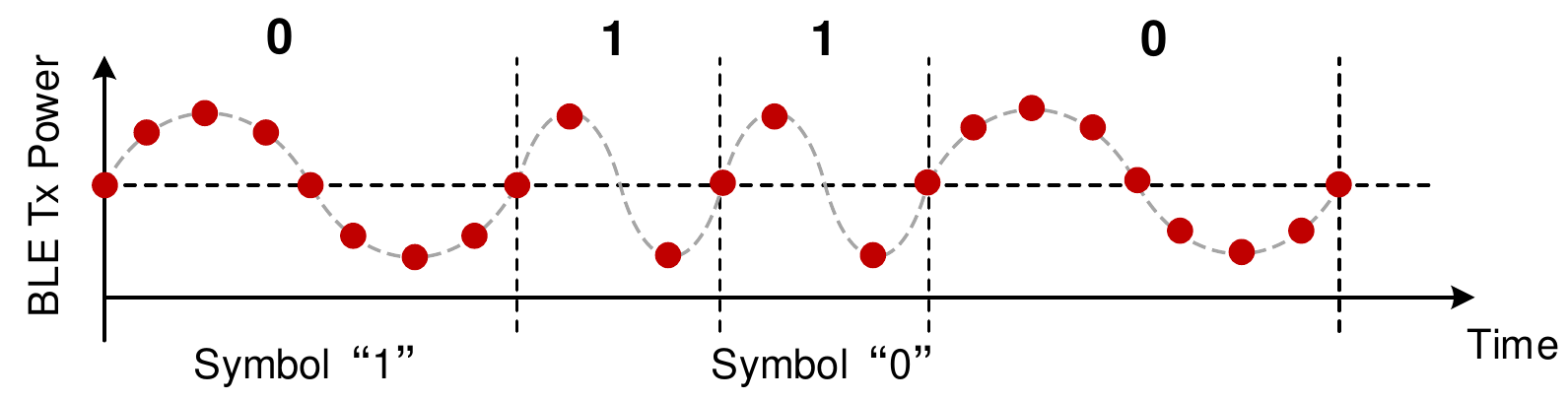}
\caption{Illustration of DAFSK in $B^2W^2$}
\label{fig:plc_b2w2_dafsk}
\end{figure}

To encode and transfer information by CSI, the $B^2W^2$ sender uses a module named DAFSK to form a sine wave by adjusting the transmission powers of the adjacent BLE packets. As shown in Fig. \ref{fig:plc_b2w2_dafsk}, discrete points on the sine wave is corresponding to the BLE packets with different transmission powers. Then the frequency of the sine wave can be changed so that the BLE data streams with different frequencies can represent different CTC symbols. For example, a sine wave with the total duration of 8 BLE packets and 4 BLE packets represents CTC symbol ``0'' and ``1'', respectively. To decode and receive information embedded in the CSI values, the $B^2W^2$ receiver extracts the CSI values that are affected by the BLE's transmission and then recovers the discrete sine waves. The $B^2W^2$ receiver demodulates the CTC symbol according to the frequency of the sine waves. Compared with FreeBee, $B^2W^2$ achieves $85$x throughput improvement by DAFSK.

\textbf{ZigFi} \cite{ZigFi} is another work depending on the side channel of CSI, which achieves CTC from a ZigBee sender to a WiFi receiver. The basic idea of ZigFi is to carefully piggy-back ZigBee packets over WiFi packets, without destroying the ongoing WiFi transmissions. In order to use the CSI sequence to enable ZigBee to WiFi CTC, some conditions need to be satisfied: (\romannumeral1) An appropriate subchannel should be selected to make ZigBee and WiFi overlap in the frequency domain. (\romannumeral2) The ZigBee packet length must be large enough to make ZigBee packets overlap with WiFi packets in the time domain. (\romannumeral3) An appropriate ZigBee power to make the CSI sequence more distinctive. The ZigFi sender transmits ZigBee packets that satisfy the above conditions and encodes the CTC symbols using the presence or absence of ZigBee packets. The ZigFi receiver receives two sets of information. It decodes packets transmitted by the ZigFi sender as a regular WiFi packet. It also collects the CSI sequence and uses the support vector machine classifier to decode the CTC data.
ZigFi achieves a throughput of $215.9bps$, which is 18x faster than FreeBee.

\begin{figure}[!tb]
\centering
\includegraphics[width=3.5in]{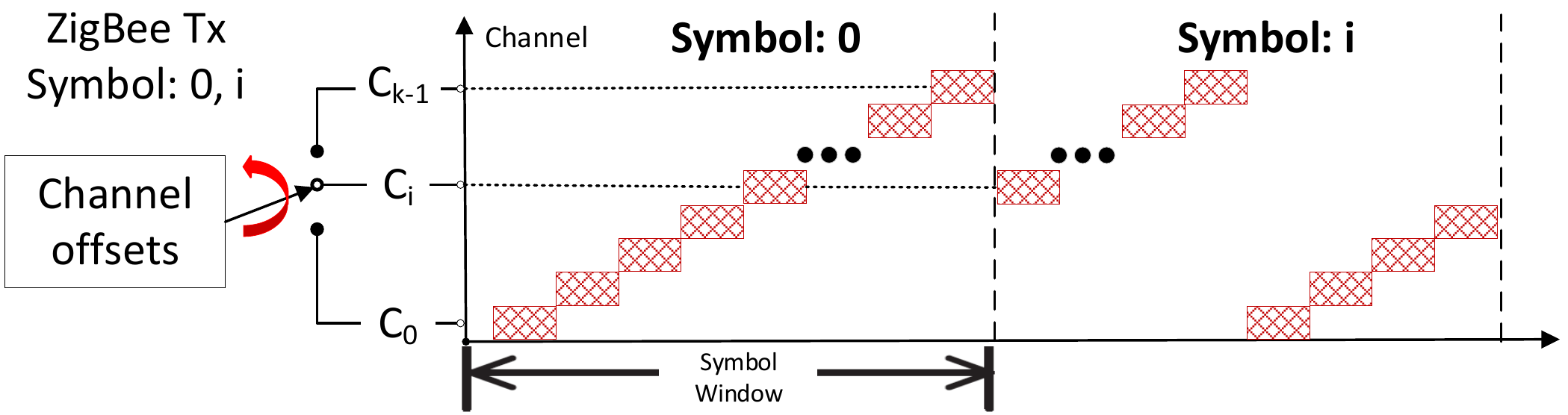}
\caption{The transmissions of ZigBee in cChirp}
\label{fig:plc_cChirp}
\end{figure}

\textbf{AdaComm} \cite{AdaComm} is also related to CSI-based CTC from ZigBee to WiFi. AdaComm aims at maintaining reliable communication performance in dynamic channel conditions. Different from previous work like ZigFi that reactively adjusts the CTC scheme at the sender, AdaComm improves reliability by using an online learning mechanism at the receiver side. The evaluation results demonstrate that AdaComm can significantly reduce the symbol error rate (SER) by $72.9\%$ and $49.2\%$, respectively, compared with the existing approaches.

\textbf{cChirp} \cite{cChirp} extends the communication range of CSI-based CTC from ZigBee to WiFi. Due to the asymmetric bandwidth and transmission power, the existing CTC from the low-power and narrow-band technology to the high-power and wide-band technology suffers from serious symbol distortions and has a limited range. Inspired by chirp spread spectrum (CSS) modulation in LoRa, cChirp proposes that the ZigBee sender can use transmissions on multiple ZigBee channels to construct chirps in the WiFi CSI matrix in Fig. \ref{fig:plc_cChirp}. This method constructs a stable and distinguishable pattern to convey symbols, improves the decoding sensitivity at the WiFi receiver, and achieves a $60m$ communication range from ZigBee to WiFi with a goodput of about $90bps$.

\textbf{DopplerFi} \cite{DopplerFi} explores how to build a CTC channel between BLE and WiFi without modifying the MAC-related configurations such as transmission power or time. As for transmission from BLE to WiFi, DopplerFi is transparent to upper layers and achieves CSI-based CTC from BLE to WiFi with a throughput of $1.59 Kbps$. DopplerFi takes advantage of the fact that the current designs of BLE can tolerate and compensate 150 KHz central frequency offset  (CFO). Particularly, BLE packets are shifted with $\pm 80$ KHz by CFO calibration in PHY. Such shift has little impact on adjacent channels due to the guard band protection and leaves enough space for BLE to tune carrier frequency and encode bits. It also ensures CFO recovery in legacy packet reception even in the presence of inherent CFO and Doppler effect. Since one BLE channel overlaps with multiple subcarriers in WiFi, WiFi can obtain the information from BLE by extracting the CSI values of WiFi packets. To identify different amounts of frequency shifts in BLE packets, WiFi differentiates them by analyzing the frequency correlations among adjacent CSI values.

\section{Physical level CTC}
\label{sec:physical}

\begin{figure}[!tb]
\centering
\includegraphics[width=3in]{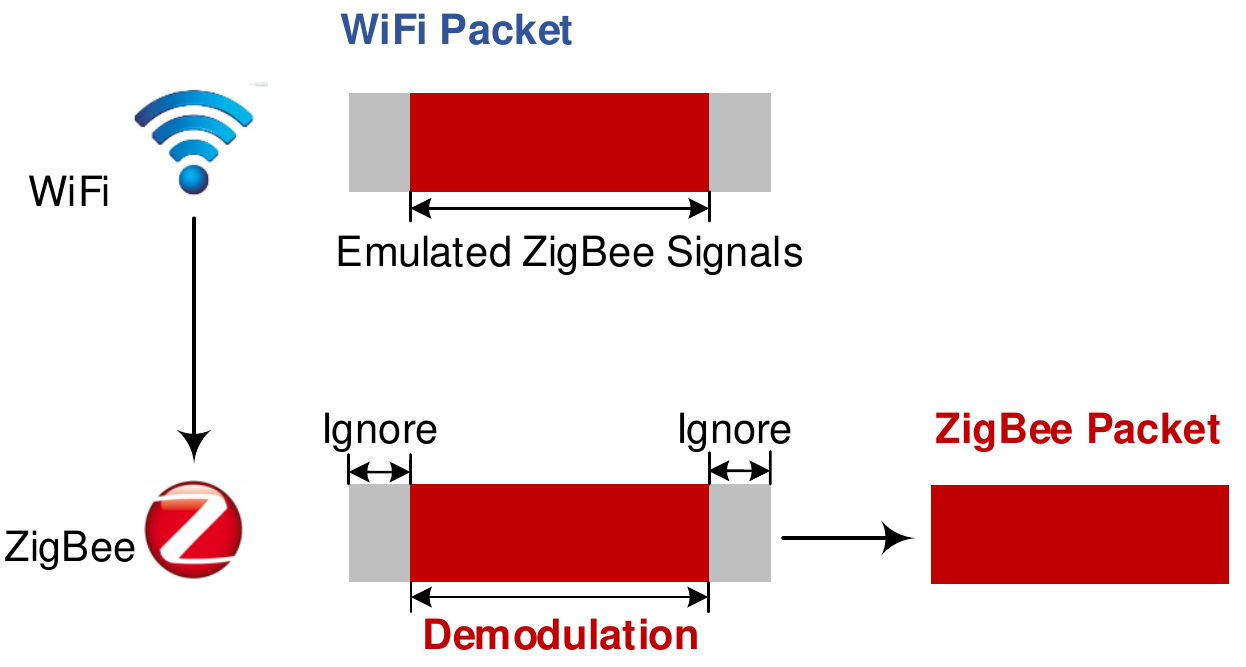}
\caption{The architecture of the WEBee}
\label{fig:webee_design}
\end{figure}

The CTC works that depends on the side channel of packet energy, packet size, packet interval, and CSI belong to the packet level CTC. The efficiency of these CTC works is bounded due to the limited throughput. First, the duration and interval of the wireless packet are in the range of milliseconds. Hence, embedding CTC symbols into the sparse wireless packets is inefficient. Second, the packet level CTC fails to fully utilize the bandwidth. Take the CTC from WiFi to ZigBee as an example. ZigBee conducts RSSI sensing within a 2MHz-bandwidth channel, while the bandwidth of WiFi is 20MHz. The signals within the rest of the WiFi bandwidth will be wasted. Due to the above limitations, the physical level CTC works arise.

According to whether the communication process is transparent to the transmitter or receiver, the physical level CTCs can be divided into three categories, namely receiver transparent CTCs, transmitter transparent CTCs and none transparent CTCs. In this section, we will introduce the representative CTC works in the three categories. Table \ref{table: existing CTCs} also includes a brief summary and comparison of them.

\subsection{Receiver Transparent CTCs}

With receiver transparent CTC, the receivers can demodulate the heterogeneous transmitters' signals without any modification. The receiver transparent CTCs mainly utilize the transmitter's signals (e.g., WiFi) to emulate the receiver's signals (e.g., ZigBee) by manipulating the transmitter's payload. According to the emulation target, these works can be divided into two categories, namely time-domain waveform-based emulation and phase-shift-sequence-based emulation. The former mainly emulates the receiver's time-domain waveform, such as WEBee \cite{WEBee} and PMC \cite{PMC}, while the latter mainly emulates the phase shift sequence of the receiver's signal, such as WIDE \cite{wide} and BlueBee \cite{bluebee}.

\begin{figure}[!tb]
\centering
\includegraphics[width=3in]{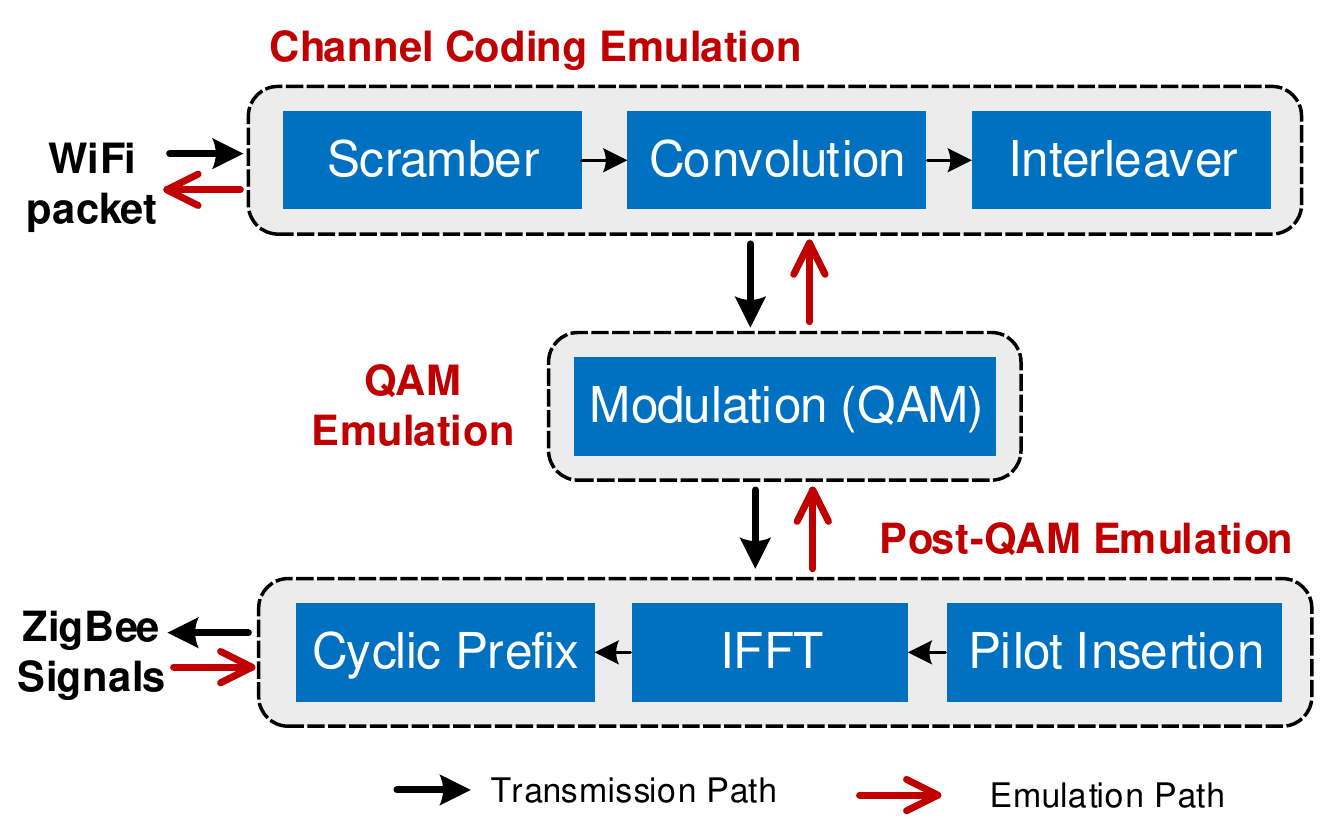}
\caption{Complete WEBee emulation procedure}
\label{fig:webee_procedure}
\end{figure}

\textbf{WEBee} \cite{WEBee} introduces a high-throughput CTC from WiFi to ZigBee via emulating the ZigBee time-domain waveform by modifying the WiFi transmitter. Fig. \ref{fig:webee_design} illustrates the architecture of WEBee. The WiFi device chooses the payload of a WiFi frame to emulate the ZigBee packet. When the ZigBee device receives signals, the WiFi header, preamble, and trailer are ignored as noise. The WiFi payload is recognized as a legitimate ZigBee packet and is decoded successfully at the ZigBee receiver. The complete WEBee emulation procedure is transparent to the ZigBee receiver and is shown in Fig. \ref{fig:webee_procedure}, which mainly consists of three parts: (\romannumeral1) Quadrature Amplitude Modulation (QAM) Emulation, (\romannumeral2) Channel Coding 
Emulation, and (\romannumeral3) Post-QAM Emulation.

\begin{figure}[!tb]
\centering
\includegraphics[width=3.2in]{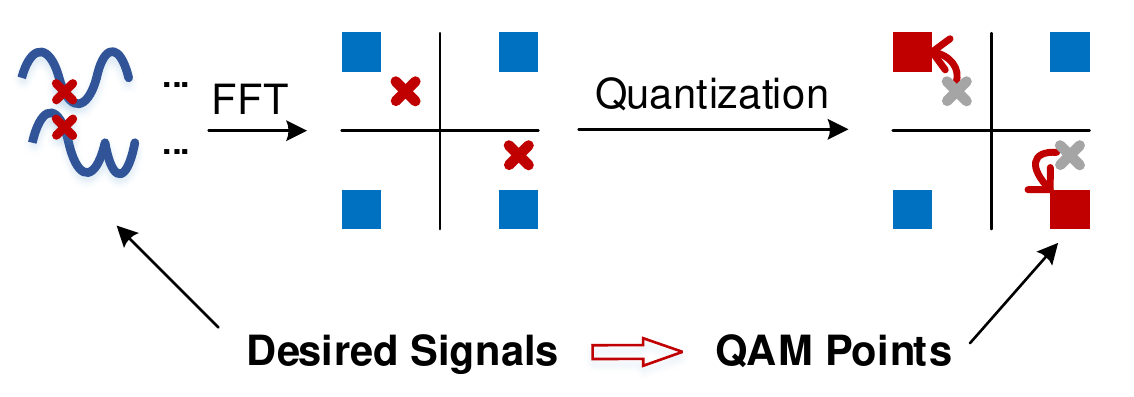}
\caption{The basic process of QAM emulation}
\label{fig:webee_fft}
\end{figure}

\emph{QAM Emulation} is the core of WEBee. As shown in Fig. \ref{fig:webee_fft}, the process of QAM selection is done in the reverse direction, where the desired ZigBee time-domain signals are sent into the fast Fourier transform (FFT) to select the corresponding QAM constellation points. Whereas the frequency components of the desired ZigBee time-domain signals may not match the WiFi QAM points perfectly, which leads to QAM quantization errors. Parseval's theorem states that the energy in the time-domain is equal to the energy in the frequency domain. That means minimizing the signal distortion in the time-domain caused by the QAM emulation errors is equal to minimizing the deviation of frequency components. Hence, the QAM emulation is an optimizing process to choose the closest $n$ QAM points in terms of the difference of FFT points between the desired signals and WiFi signals. Moreover, the Direct Sequence Spread Spectrum (DSSS) also improves the ability to tolerate errors. Specifically, a ZigBee symbol (4-bits) is mapped into a 32-chip sequence. The maximum Hamming distance between the received chip sequence and the standard chip sequence is customizable in commercial off-the-shelf devices. Hence, in WEBee, the maximum Hamming distance can be set more loosely to tolerate QAM emulation errors.

\emph{Channel Coding Emulation} is used to achieve the emulation of convolutional encoder, scramber, and interleaver. First, the convolutional encoding can be modeled as a matrix $M$ , which satisfies $M \times_{GF(2)}X=Y$. The Galois field $GF(2)$ is used to define the relationship between the source bits $X$ and the coded bits $Y$. WEBee only needs to control 7 WiFi QAM points to emulate ZigBee signals because the ZigBee channel only covers 7 WiFi subcarriers. So with 64-QAM, WEBee controls only 42 (7$\times$6) bits of $Y$ by manipulating $X$. In addition to convolutional encoder, the scrambling of WiFi is achieved by XORing the incoming source bits with the output of a 7-bit linear feedback shift register. It is easy to reverse the scrambler by XORing the scrambled bits with the same output of the shift register because the scrambler is a one-to-one mapping from the source bits to the scrambled bits. Similarly, the interleaver is also a one-to-one mapping from the coded bits to the permuted bits, and it can be reversed easily.

\emph{Post-QAM Emulation} has several challenges to be resolved. First, the duration of a ZigBee symbol is four times that of WiFi. Therefore, a complete ZigBee symbol has to be segmented before emulated by four WiFi symbols and such segmentation introduces boundary errors. Second, the cyclic prefix (CP) will also cause errors. The 0.8~$\mu$s-CP means that the front segment and the end segment of WiFi signals are the same, introducing additional error for ZigBee.
Due to the inherent difference between WiFi and ZigBee, the signal distortion cannot be avoided completely during the emulation. WEBee also proposes repeated transmission and forward error correction (FEC) for reliability.

\begin{figure}[!tb]
\centering
\includegraphics[width=3.3in]{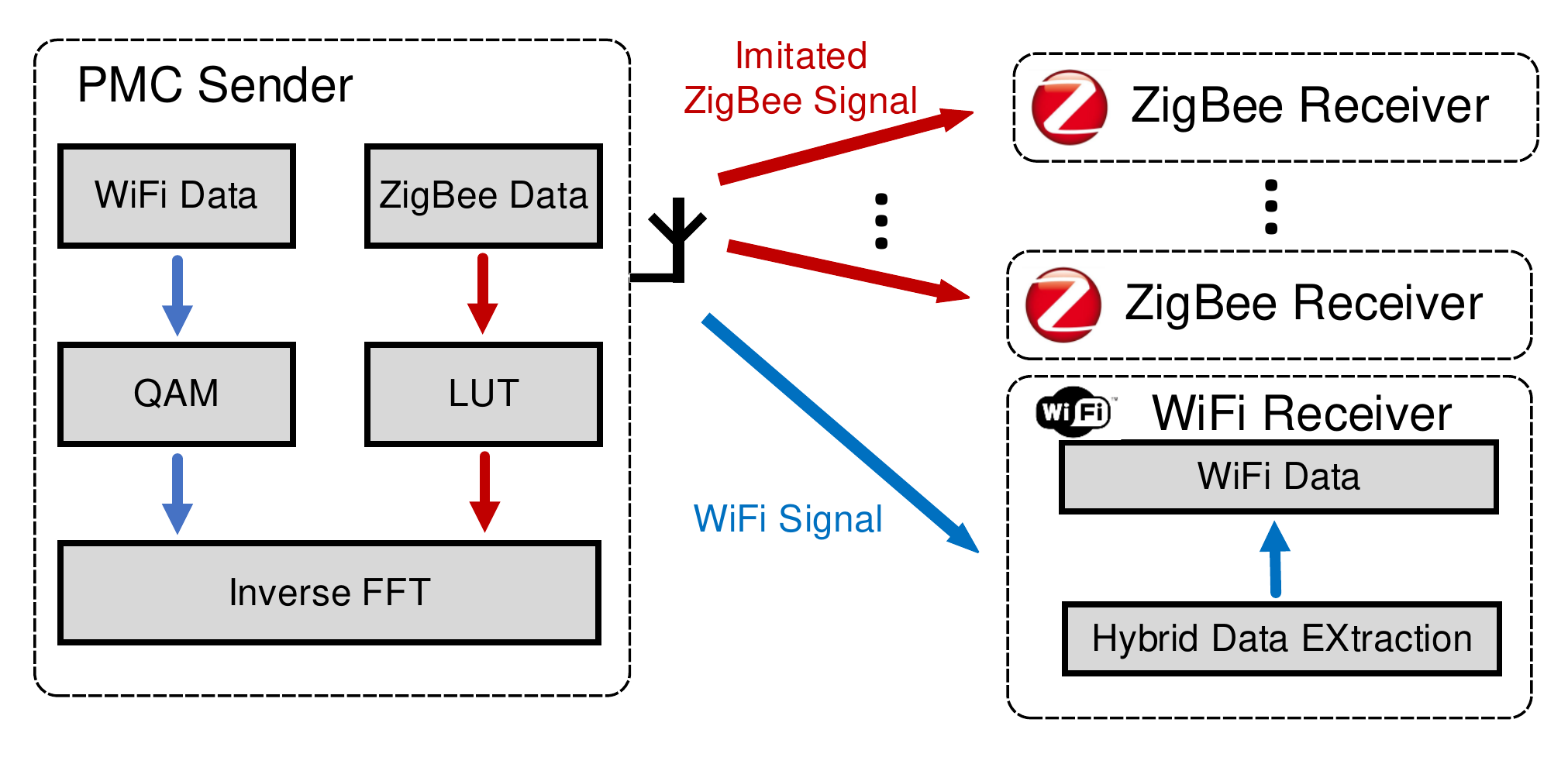}
\caption{The system overview of the PMC}
\label{fig:pmc}
\end{figure}

\textbf{PMC} \cite{PMC} is another waveform-based CTC from WiFi to ZigBee. Different from WEBee, PMC only uses the signals of the overlapping WiFi subcarriers to emulate ZigBee signals. The other subcarriers still transmit WiFi signals. The system overview is shown in Fig. \ref{fig:pmc}. Specifically, PMC firstly develops an offline search algorithm, which can map the desired ZigBee signals to WiFi QAM-modulated signals. This search algorithm iteratively finds the QAM phase states that are most similar to the ZigBee offset quadrature phase shift keying (OQPSK) signals from all possible QAM phase states in the overlapping WiFi subcarriers. It should be noted that it is different from WEBee \cite{WEBee} as it can choose other QAM points besides the original WiFi constellation points. This mapping relationship is stored in a look-up table (LUT). Then WiFi sender goes through the LUT table to get the QAM states corresponding to the ZigBee signals. The signals in other non-overlapping subcarriers are the traditional WiFi signals. The hybrid ZigBee and WiFi signals are sent after the inverse FFT module. In this way, the hybrid signals can be decoded by the WiFi receiver and ZigBee receiver, respectively. It should be noted that the hybrid signals can be directly received and decoded by the ZigBee receiver without any modification, while the WiFi receiver needs to be modified at the link layer to extract the WiFi data from the hybrid signals.

In addition to the person-area network (e.g., WiFi, Zigbee, and Bluetooth), some works focus on the wide-area network (e.g., LTE and Multefire). \textbf{LTE2B} \cite{lte2b} is a representative CTC work that delivers information from LTE to ZigBee. 
Some works attempt to combine the physical level CTC with backscatter. \textbf{Passive-ZigBee} \cite{passivezigbee} utilizes a low power backscatter radio to transform a WiFi signal into a ZigBee packet.

\begin{figure}[!tb]
\centering
\includegraphics[width=3.2in]{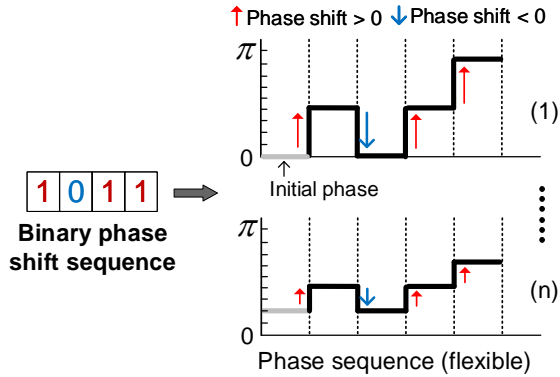}
\caption{The process of digital emulation}
\label{fig:wide}
\end{figure}


On the one hand, in spite of the progress in time-domain waveform emulation, an important fact is often overlooked: the emulated signals from the sender cannot perfectly match the desired signals of the receiver due to the difference in communication standards and the hardware restrictions. There is more or less Hamming distance between the emulated and the desired signals, incurring emulation errors. 
On the other hand, the receiver transparent CTC receiver (ZigBee, BLE, etc.) uses the phase shift rather than the phase itself or time-domain waveform to decode signals. Specifically, the receiver outputs ``1'' if the phase shift between two consecutive samples is bigger than 0 and otherwise outputs ``0''.

Based on the above finding, \textbf{WIDE} \cite{wide} achieves CTC based on the method of digital emulation to reduce emulation errors. Instead of emulating the original time-domain waveform of the receiver, the sender emulates the phase shifts associated with the desired signals. The process of digital emulation is shown in Fig. \ref{fig:wide}. Given the desired data bits of the receiver, the sender calculates the signs of phase shifts. The positive and negative phase shifts represent the bit ``1'' and ``0'', respectively. The sender generates a ladder-shaped phase sequence that matches the signs of phase shifts. The duration of each phase value is equal to the decoding period of the receiver. The ladder-shaped phase sequence corresponds to a waveform, which is then emulated by using the time-domain waveform emulation.

Compared with the time-domain waveform emulation, digital emulation is more flexible and robust.  As shown in Fig. \ref{fig:wide}, the phase shift sequence of the desired data bits at the receiver side is not unique, because the receiver decodes signals according to the sign of the phase shift rather than the specific phase shift value. For example, the phase shift value $+\frac{\pi}{4}$ and $+\frac{\pi}{2}$ can both be decoded as ``1'' due to the positive sign of phase shift. The errors at the WiFi sender when emulating different phase shift sequences are different. Therefore, we have the opportunity to reduce the emulation errors by selecting an appropriate phase shift sequence for emulation.

\begin{figure}[!tb]
\centering
\subfigure{\label{fig:bluebee_zigbee}}\addtocounter{subfigure}{-2}
\subfigure[ZigBee signal with two chips ``11'' ]{\subfigure
{\includegraphics[width=0.45\linewidth]{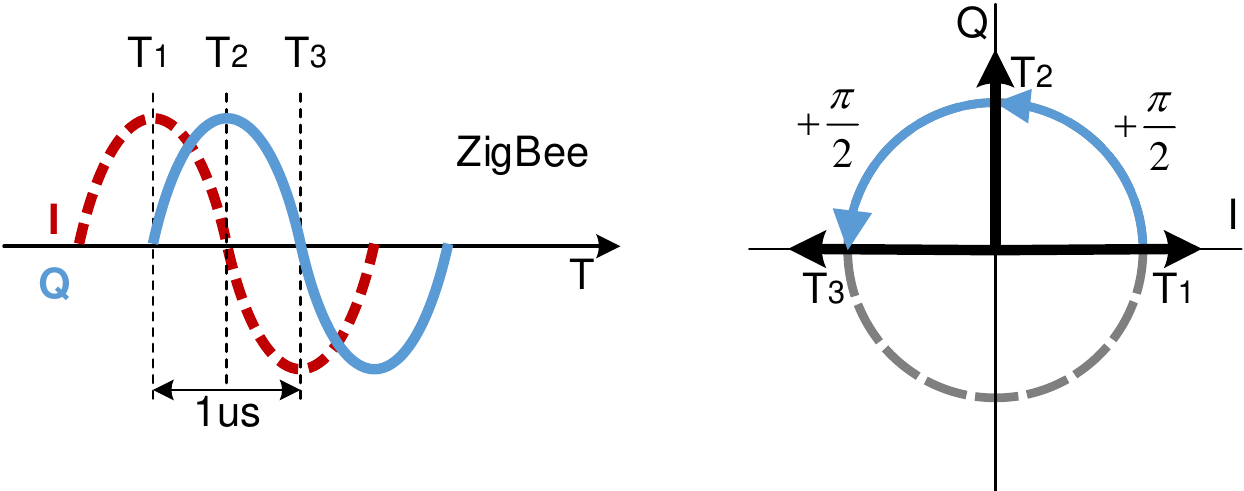}}}
\subfigure{\label{fig:bluebee_ble}}\addtocounter{subfigure}{-2}
\subfigure[Emulation of (a) by BLE]{\subfigure
{\includegraphics[width=0.45\linewidth]{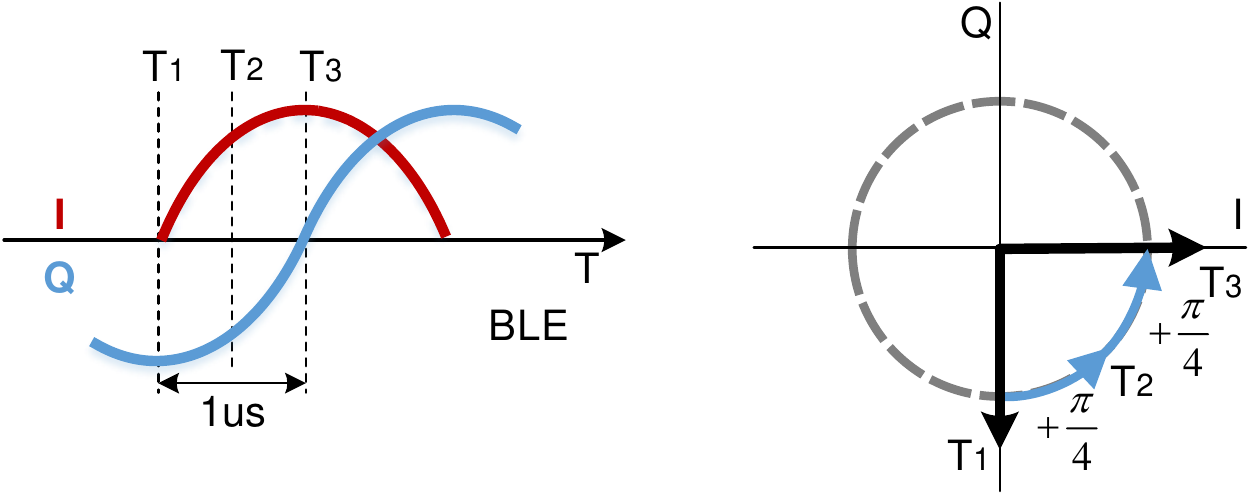}}}
\caption{The emulation process of BlueBee}
\end{figure}

\textbf{BlueBee} \cite{bluebee} proposes a CTC from BLE to ZigBee by emulating legitimate ZigBee packets using BLE packets. It is also based on phase shift sequence emulation. The feasibility of emulating ZigBee packets using BLE packets relies on two key technique insights. First, the modulation techniques of ZigBee and BLE are similar. ZigBee's OQPSK and BLE's GFSK both leverage the phase shift between consecutive samples to indicate symbols. Second, the demodulation of Zigbee only considers the sign of the phase shift (``+'' or ``-'') instead of a particular phase shift value, which offers great flexibility in emulation. Fig. \ref{fig:bluebee_zigbee} depicts the ZigBee signal containing chips ``11''. The phase shifts from $T_1$ to $T_2$ and from $T_2$ to $T_3$ are both $\frac {\pi}{2}$. Fig. \ref{fig:bluebee_ble} shows the BLE signal, which is the emulation of Fig. \ref{fig:bluebee_zigbee}. The bandwidth of BLE is half of the bandwidth of ZigBee, which means the sampling rate of BLE is also half of ZigBee. When the BLE signal is fed into the ZigBee receiver, the ZigBee receives samples at $T_1$, $T_2$, and $T_3$. The phase shifts from $T_1$ to $T_2$ and from $T_2$ to $T_3$ are both $\frac{\pi}{4}$. Since the signs of these two phase shifts are positive, the ZigBee chips can be successfully decoded as ``11''. In this case, the ZigBee receiver can decode the BLE signal segment as ``11'' or ``00''. Consider that there are ``10'' and ``01'' in the DSSS sequence of the ZigBee symbols. The BLE signal is optimally designed such that the inevitable error is minimized and kept under the tolerance of the ZigBee's OQPSK/DSSS demodulator.

\subsection{Transmitter Transparent CTCs}

Different from the receiver transparent CTCs that utilize the transmitter's strong capability to realize the communication from high-end transmitters to low-end receivers, the transmitter transparent CTCs make full use of the receiver's capability to realize the communication from low-end transmitters to high-end receivers without any modification of transmitters. Now we introduce three transmitter transparent CTC works. The first two of them observe the pattern of the transmitter's signal at the receiver to achieve cross-decoding, while the third one utilizes the strong computing capability of the WiFi receiver to reconstruct the transmitter's signal.

\begin{figure}[!tb]
\centering
\includegraphics[width=3.2in]{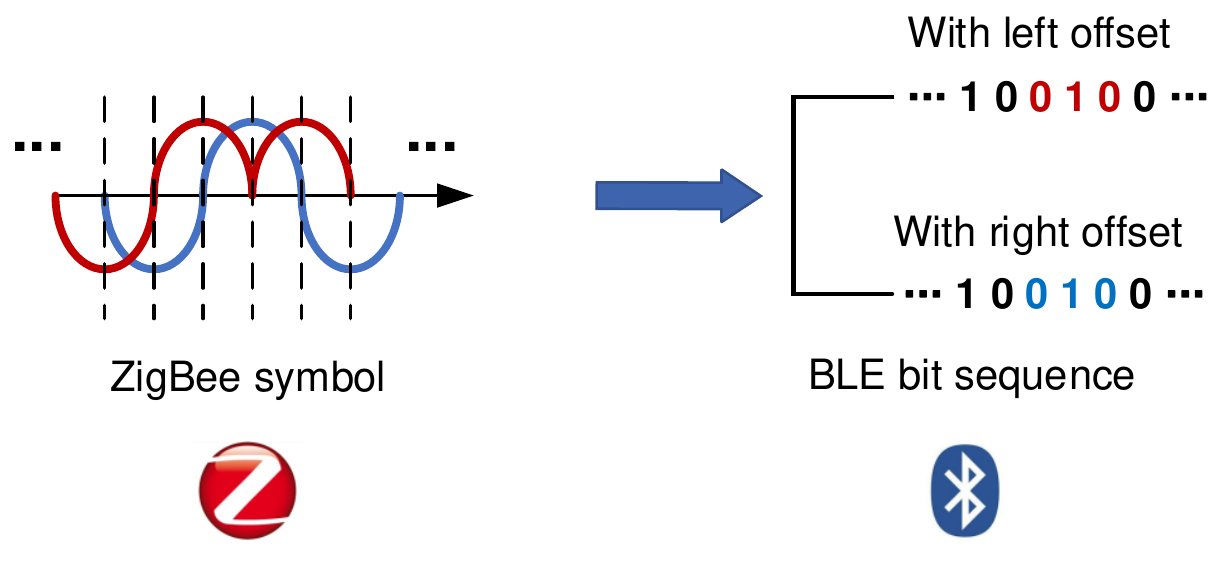}
\caption{Two different bit sequences obtained by the BLE receiver for ZigBee symbol}
\label{fig:xbee}
\end{figure}

\textbf{XBee} \cite{xbee} is a physical-level CTC from ZigBee to BLE. This work proposes the method of \textit{cross-decoding}, which interprets a ZigBee frame by observing the bit pattern obtained at the BLE receiver. Cross-decoding is inspired by the following two technical insights. First, Both the ZigBee receiver and the BLE receiver utilize the phase shift to decode their signals. Second, the phase shifts at the BLE receiver are quantized, and only the sign of phase shifts are used. We illustrate the method of cross-decoding with the example shown in Fig. \ref{fig:xbee}. A ZigBee symbol lasts 16$\mu$s. Considering the sampling rate of BLE is 1$MHz$, the above ZigBee signals can be demodulated as 8 BLE bits based on the sign of phase shifts. Since the sampling rate of BLE is half of the sampling rate of ZigBee, whether the samples have a left offset or a right offset determines the final decoding result. The BLE decoding bit sequence has some determined bits and some undetermined bits. According to the demodulated BLE bits, the BLE receiver can infer the ZigBee symbols.

\begin{figure}[!tb]
\centering
\includegraphics[width=3.2in]{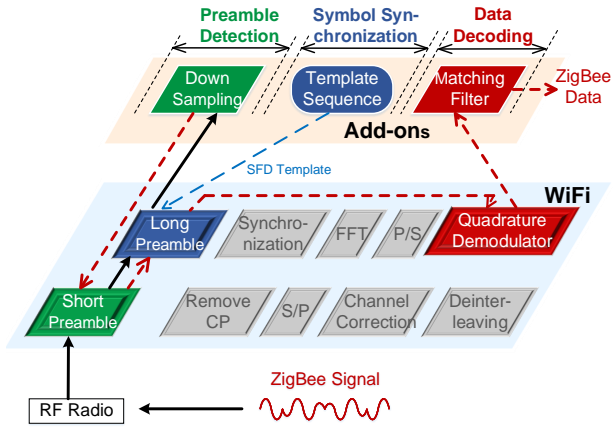}
\caption{The framework of LEGO-Fi \cite{legofi}}
\label{fig:legofi}
\end{figure}

\textbf{LEGO-Fi} \cite{legofi} is another transmitter transparent CTC, which delivers information from ZigBee to WiFi. LEGO-Fi reuses the standard WiFi modules for the ZigBee reception and proposes a concept named \textit{cross-demapping}. As shown in Fig. \ref{fig:legofi}, the authors reuse the WiFi short preamble detection module, the WiFi long preamble detection module, and the quadrature demodulation module to decode ZigBee symbols. First, the received signals after the process of downsampling are fed to the WiFi short preamble detection module. Second, if the periodic ZigBee preamble is detected, we reuse the WiFi long preamble detection module to conduct symbol synchronization to segment each ZigBee symbol. During this process, the start of frame delimiter (SFD) template of ZigBee is fed into the WiFi long preamble detection module to locate the SFD. Third, these received signals are forwarded to the quadrature demodulator to calculate the corresponding phase shift sequence. Finally, LEGO-Fi uses a matching filter to distinguish different ZigBee symbols and accomplish CTC from ZigBee to WiFi.

\begin{figure}[!tb]
\centering
\includegraphics[width=3in]{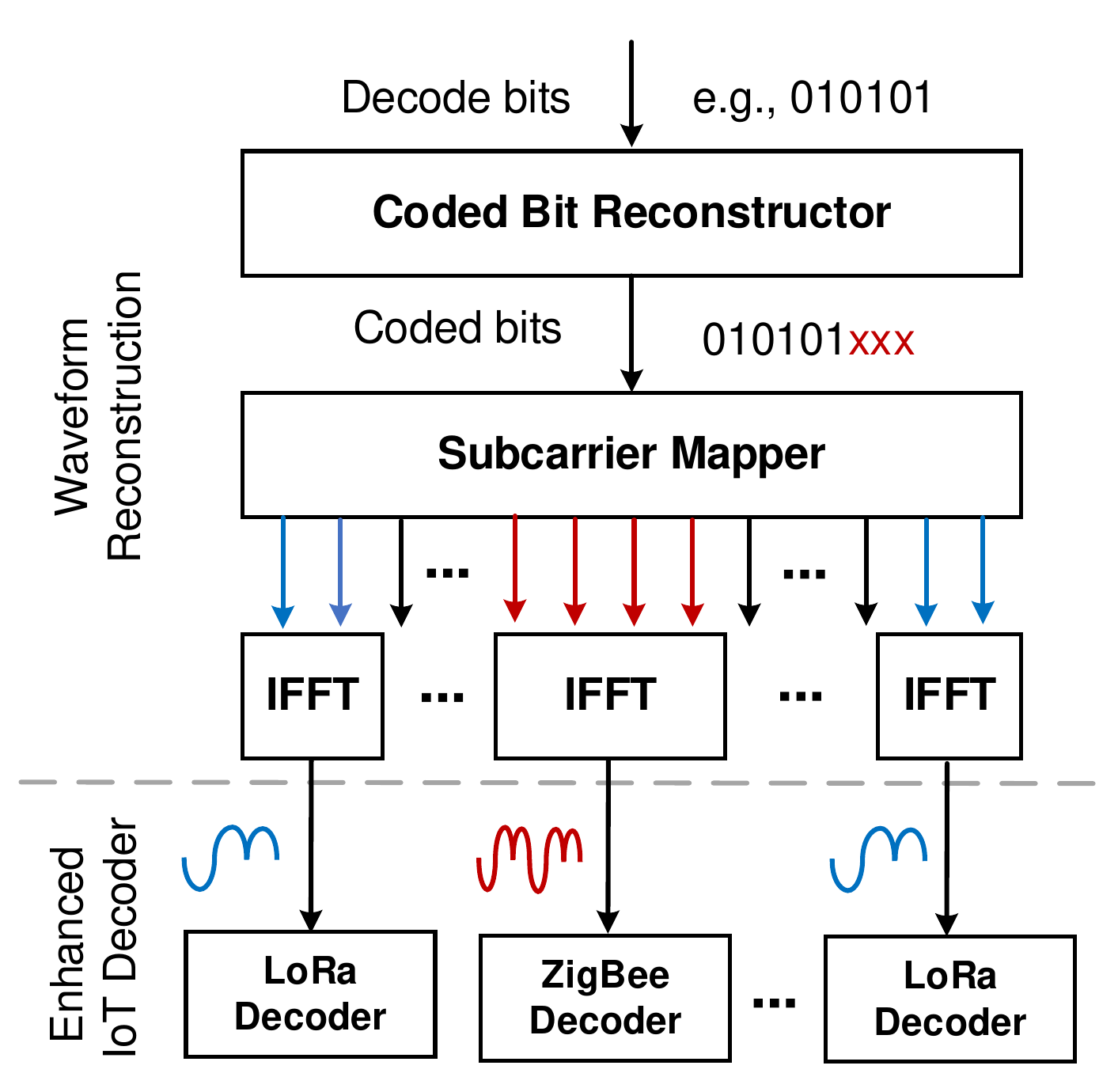}
\caption{The architecture of XFi}
\label{fig:xfi}
\end{figure}

\textbf{XFi} \cite{xfi} enables mobile devices to directly and simultaneously collect data from diverse IoT devices by commodity WiFi radio. The key insight is that the IoT data can be captured by the WiFi receiver and retained when the IoT frame collides with a WiFi transmission. XFi obtains the collided IoT data by analyzing the decoded WiFi payload. The detailed procedures of XFi are as follows: (\romannumeral1) reconstruct the waveform of hitchhiking IoT data. (\romannumeral2) decode the reconstructed IoT waveform. Fig. \ref{fig:xfi} shows the architecture of XFi. The coded bits are recovered from decode bits by the coded bit reconstructor. The coded bits are mapped to the subcarriers and the IoT waveform can be reconstructed by performing IFFT. A robust decoding algorithm is used to decode the IoT data with these reconstructed waveforms.  While the channel decoder of WiFi adopts FEC and attempts to eliminate the hitchhiking IoT signal as interference in the decoded output, the author observes that the decoder almost keeps the corrupted coded bits intact, especially when coded bits are severely disturbed by the IoT signal. So the coded bits can be approximated with decoded bits. On the other hand, with the \textit{parity removal} and \textit{CP removal}, nearly a third of IoT waveforms are erased by WiFi hardware and cannot be reconstructed. In this case, XFi customizes an enhanced IoT decoder to provide robust decoding with the symbol-level and chip-level redundancy of IoT signals. 

\subsection{None Transparent CTCs}

In addition to the above two types of CTCs, another type of CTC modifies both the transmitter and the receiver. Part of them are to enhance the robustness of CTC, such as TwinBee \cite{TwinBee}, LongBee \cite{LoneBee} and SymBee \cite{symbee}. Other works are to achieve parallel communication between the different wireless protocols, such as Chiron \cite{chiron} and PIC \cite{pic}. We introduce these representative works below.

\textbf{TwinBee} \cite{TwinBee} is a representative none transparent CTC work to enhance the robustness of CTC, which is proposed to recover errors introduced by imperfect signal emulation of WEBee. The author analyzes the reasons for these errors and conducts several experiments to explore the chip error patterns. The received 32-chip sequence of ZigBee, whose chip errors are located in the middle and both ends of the chip sequence, is regarded as the error-prone chip. The rest of the chips are regarded as normal chips. Since those chip errors have distinguishable patterns, TwinBee designs a specific chip-combining coding method to recover the errors in error-prone chips.

The cyclic-shift feature of ZigBee chip sequence ensures the feasibility of chip-combining coding. As we know, a 4-bit ZigBee symbol is mapped into a 32-chip sequence, and there are totally 16 different chip sequences for symbol ``0'' to symbol ``15''. The chip sequence ``$m+2$'' is the right-cyclic-shifted by 4 chips from the chip sequence ``$m$''.

The basic idea of the chip-combining coding is leveraging the cyclic-shift feature of ZigBee chip sequences to move the error-prone chips away. We suppose the length of error-prone chips is 8 chips. After the chip sequence ``$m+2$'' is shifted by 8 chips, the error-prone chips' position is exactly complementary to the position of symbol ``$m$''. Combining these two emulated symbols only with their normal chips will recover the original symbol with all normal chips.

\begin{figure}[!tb]
\centering
\includegraphics[width=3.5in]{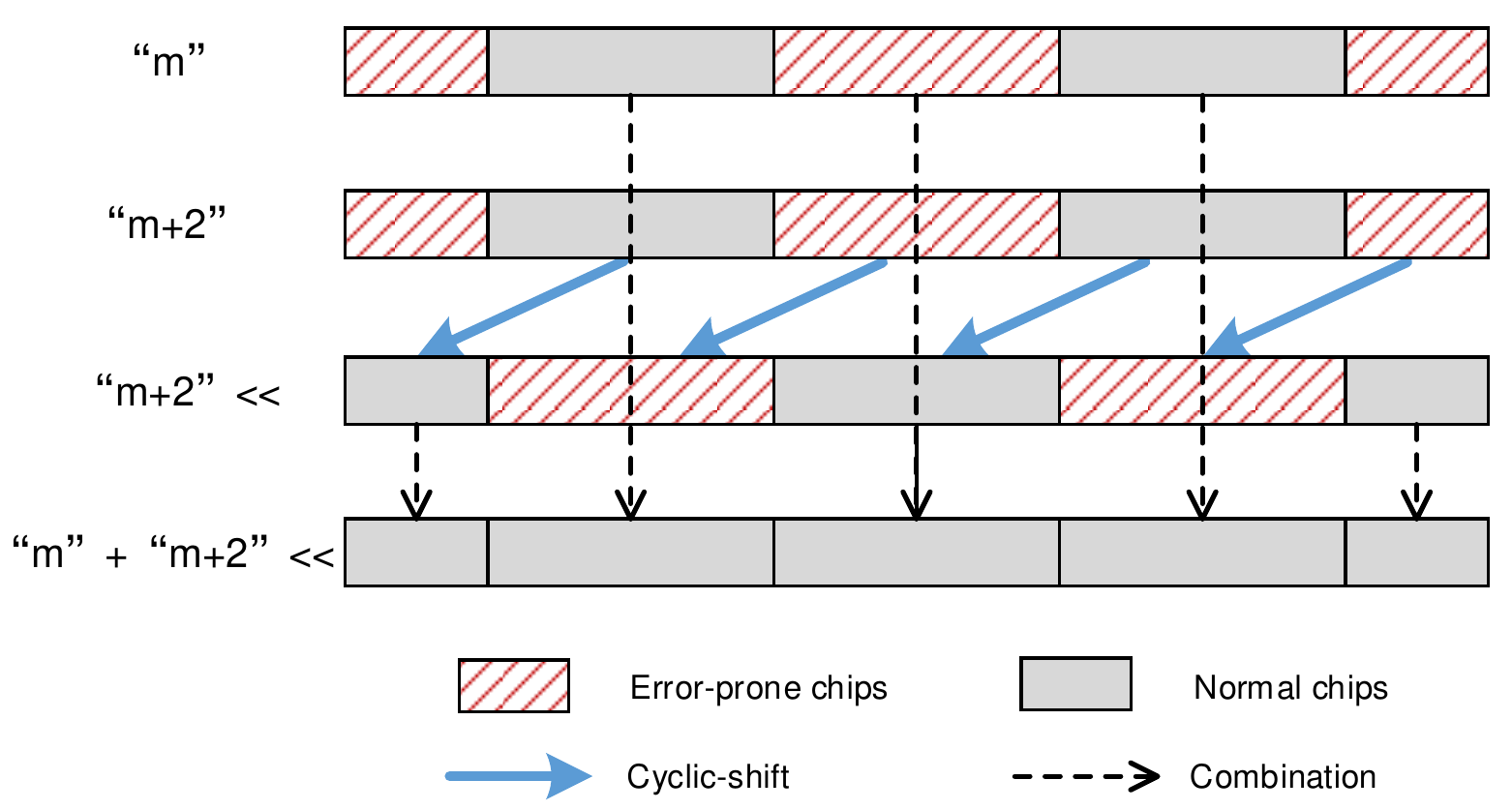}
\caption{The diagram of chip-combining coding in TwinBee}
\label{fig:twinbee}
\end{figure}

The diagram of chip-combining coding is shown in Fig. \ref{fig:twinbee}. An original symbol ``$m$'' is to be transmitted. The TwinBee sender firstly selects a twin symbol ``$m+2$'' whose chip sequence is right-cycle-shifted by 8 chips from the original symbol. Then the TwinBee sender combines these twin symbols into one byte and transmits via emulation. The TwinBee receiver left-cyclic-shifts the received chip sequence of symbol ``$m+2$'' by 8 chips, denoted as ``$m+2<<$''. The positions of error-prone chips of ``$m+2<<$'' are different from the original symbol $m$. In addition, due to the cyclic-shift feature of the ZigBee chip sequence, the chip sequence of ``$m+2<<$'' is equal to the symbol ``$m$'' in theory. The TwinBee receiver combines the normal chips of these two chip sequences ``$m+2<<$'' and ``$m$'' together. In this way, the TwinBee receiver gets the chip sequence of the original symbol ``$m$'' with all normal chips, and the error-prone chips can be recovered successfully.

\textbf{LongBee} \cite{LoneBee} is another improved CTC work of WEBee. LongBee extends the communication range of CTC to support long-range IoT applications. In terms of signal emulation, LongBee works similarly to WEBee. Moreover, LongBee combines the high transmission power of WiFi and the fine receiving sensitivity of ZigBee together to increase the CTC communication range significantly. 
\textbf{SymBee} \cite{symbee} is a symbol-level CTC from ZigBee to WiFi, which is built on the insight of cross-observability on ZigBee-WiFi physical layer. The ZigBee sender transmits specific symbols, and these symbols yield unique and easily detectable patterns when cross-observed at the WiFi receiver. SymBee elaborately selects optimal combinations of ZigBee symbols to achieve two goals. First, these symbols yield the longest stable patterns that maximize the detection under noise and interference. Second, the difference between different combinations of elected symbols used to represent different CTC symbols is maximally distinct.
\textbf{Chiron} \cite{chiron} is a representative none transparent CTC work to achieve parallel communication, as it designs a Chiron receiver and a Chiron sender to enable parallelly transmitting (or receiving) WiFi data and Zigbee data to (or from) commodity WiFi and ZigBee devices.

\textbf{PIC} \cite{pic} design a new gateway to achieve the parallel inclusive bi-directional transmission of both WiFi and BLE data simultaneously. The core of the PIC's design is to generate a frame that contains both WiFi and BLE data which can be demodulated by both WiFi and BLE devices, leveraging the unique modulation schemes of WiFi and BLE.
\textbf{Symphony} \cite{symphony} achieves CTC from both ZigBee and BLE to LoRa.  The key ideas of this approach are two techniques: (\romannumeral1) ZigBee and BLE can both generate several specific signals by controlling the payload. (\romannumeral2) LoRa demodulation mechanism based on FFT can be used to detect the specific signals from ZigBee and BLE.

\section{Overall Comparison of CTC}
\label{sec:overhead}

In this section, we compare CTC techniques in terms of throughput, reliability, concurrency, hardware modification, and complexity overhead, as shown in Table~\ref{table: existing CTCs}. 

\subsection{Performance Comparison}

First, the data rate of packet level CTC is lower than that of physical level CTC. For example, the data rate of packet level CTC is usually no more than a few Kbps, while the data rate of physical level CTC can be up to hundreds of Kbps. That is expected because the physical level CTC uses more fine-grained time-domain signal or phase information to transmit CTC messages, while the packet level CTC uses coarse-grained packet features. A packet can only carry a few bits of CTC messages. 

Second, the packet level CTC is usually more reliable than the physical level CTC. Specifically, the packet level CTCs have stronger anti-interference and anti-noise ability than the physical level CTCs.

Third, parallel CTC is often desired can be realized by either a packet level CTC or physical level CTC. For example, the packet level CTC named DCTC~\cite{dctc} and the physical level CTC named WEBee~\cite{WEBee} support four-channel concurrent CTC links.

\subsection{Complexity Comparison}

First, the complexity and cost of the packet level CTC are relatively low.
For the RSSI-based packet level CTC and the CSI-based packet level CTC, the transmitter sometimes needs to inject extra data packets to construct distinguished RSSI and CSI patterns.
At the transmitter, the packet level CTC has no requirement for the data payload and there is no modification on the hardware and the MAC protocol. 
At the receiver, there is only an add-on demodulation algorithm based on the received RSSI sequence and CSI sequence.
The demodulation algorithm of RSSI classification is simple and the computational complexity is $O(n)$, where $n$ is the number of samples.
Whereas, the demodulation complexity of the CSI-based packet level CTC depends on the classification algorithm. Different CSI classification algorithms, such as variance detection, SVM classifier and CNN classifier, have different complexity and cost.

Second, the complexity of the physical level CTC is higher than that of the packet level CTC. We present brief comparison of them as follows.

(1) Receiver-transparent physical level CTC: In order to emulate the desired signal or phase of the receiver, the transmitter needs to manipulate the packet payload. The emulation process is the reverse-engineering of the modulation process. FFT operation is the most time-consuming operation in the process of signal emulation, and its computational complexity is $O(lgn)$ where $n$ is the length of the desired packet. There is no modification on the modulation/demodulation algorithm, transmitting/receiving parameters, hardware, or the MAC protocol at the transmitter and the receiver. 

(2) Transmitter-transparent physical level CTC: In order to demodulate the signal from the transmitter without any modification, the receiver reuses and rewires the existing processing modules to obtain a mapping table between the identified patterns and the CTC messages. The receiver only needs to update the demodulation algorithm without the modification on hardware or the MAC protocol. The transmitter also does not require any modification and transmits the data packet directly.

(3) None-transparent physical level CTC. In order to improve the reception radio of the emulated packet, extend the communication distance of CTC, and support CTC links with higher concurrency, the none-transparent physical level CTC leverages more complex operations such as bandwidth conversion and segment stitching. Although there is no modification on the hardware and MAC protocol, the transmitter needs to modify the packet payload and the transmitting parameters (bandwidth, power, and interval). The receiver employs more complex demodulation algorithm at the software layer.

\section{Upper layer CTC}
\label{sec:upperlayer}

As mentioned above, packet level CTC and physical level CTC form the physical layer foundation of direct communication among heterogeneous devices. However, A seamless CTC network requires the construction of upper network layers. Simply applying upper layer protocols of homogeneous devices is not optimal. The existence of heterogeneous devices in the network brings opportunities as well as challenges. For example, a heterogeneous network may provide a better routing solution for the network layer. But different signal strength of these devices may lead to a cross-technology hidden terminal problem in the MAC layer. Meanwhile, channel quality estimation and acknowledgment are also issues that require resolving. This section introduces some existing works on building upper layer protocols of CTC network, as shown in Table \ref{table:upperlayer}.

\begin{figure}[!tb]
\centering
\includegraphics[width=3.6in]{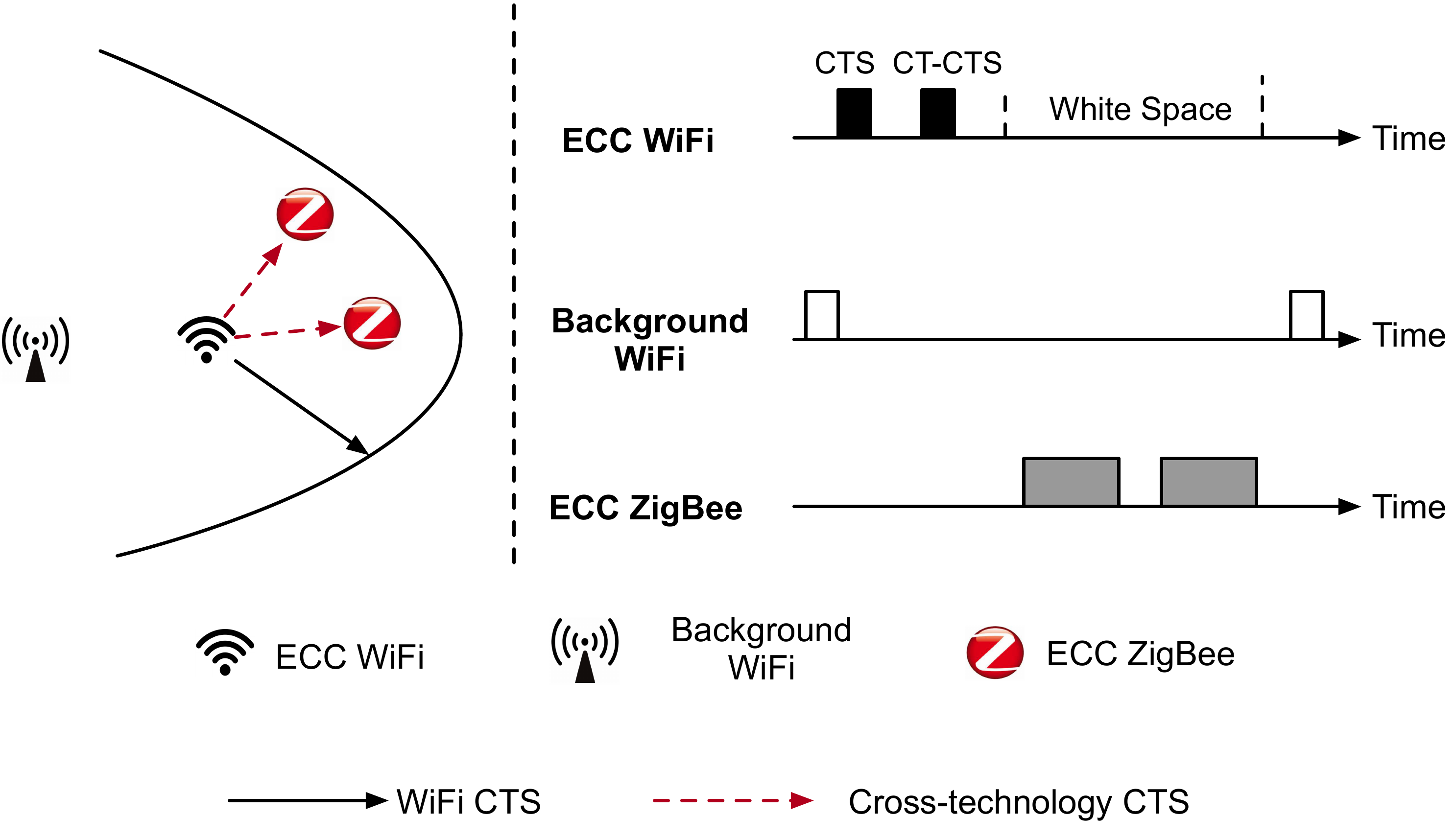}
\caption{The overview of ECC}
\label{fig:ecc}
\end{figure}

\textbf{ECC} \cite{ecc} is a representative work that achieves explicit channel coordination on top of CTC from WiFi to ZigBee, as shown in Fig.21. It ensures ZigBee communication under CTI without disrupting WiFi operation. Specifically, ECC first aggregates scattered WiFi white spaces by adopting clear to send (CTS) in the WiFi standard. Then, the WiFi sender broadcasts the cross-technology CTS via CTC, notifying the ZigBee device to wake up and access the spectrum. During the whole communication window, background WiFi devices remain silent due to CTS. In this way, ECC protects the ZigBee transmission from the interference of WiFi and improves the spectrum efficiency by recycling white spaces.

\begin{table*}[]
\centering
\caption{Upper layer works of CTC}
\begin{tabular}{|l|l|l|l|}
\hline
\textbf{Name} & \textbf{Layer} & \textbf{Design Target}  & \textbf{Technology}       \\ \hline
ECC \cite{ecc}           & MAC Layer      & Channel Coordination    & WiFi -\textgreater ZigBee \\ \hline
ECT \cite{ect}           & Network Layer  & Data Forwarding         & ZigBee -\textgreater WiFi \\ \hline
NetCTC \cite{netctc}       & Network Layer  & ACK                     & ZigBee -\textgreater WiFi \\ \hline
CRF \cite{crf}           & Network Layer  & Routing and Flooding    & ZigBee -\textgreater WiFi \\ \hline
C-LQI \cite{clqi}         & Link Layer     & Link Quality Estimation & WiFi -\textgreater ZigBee \\ \hline
XMIMO \cite{bib:X_MIMO}         & Link Layer     & Link Quality Estimation & ZigBee -\textgreater WiFi \\ \hline
\end{tabular}
\label{table:upperlayer}
\end{table*}
 
\begin{figure}[!tb]
\centering
\includegraphics[width=3.3in]{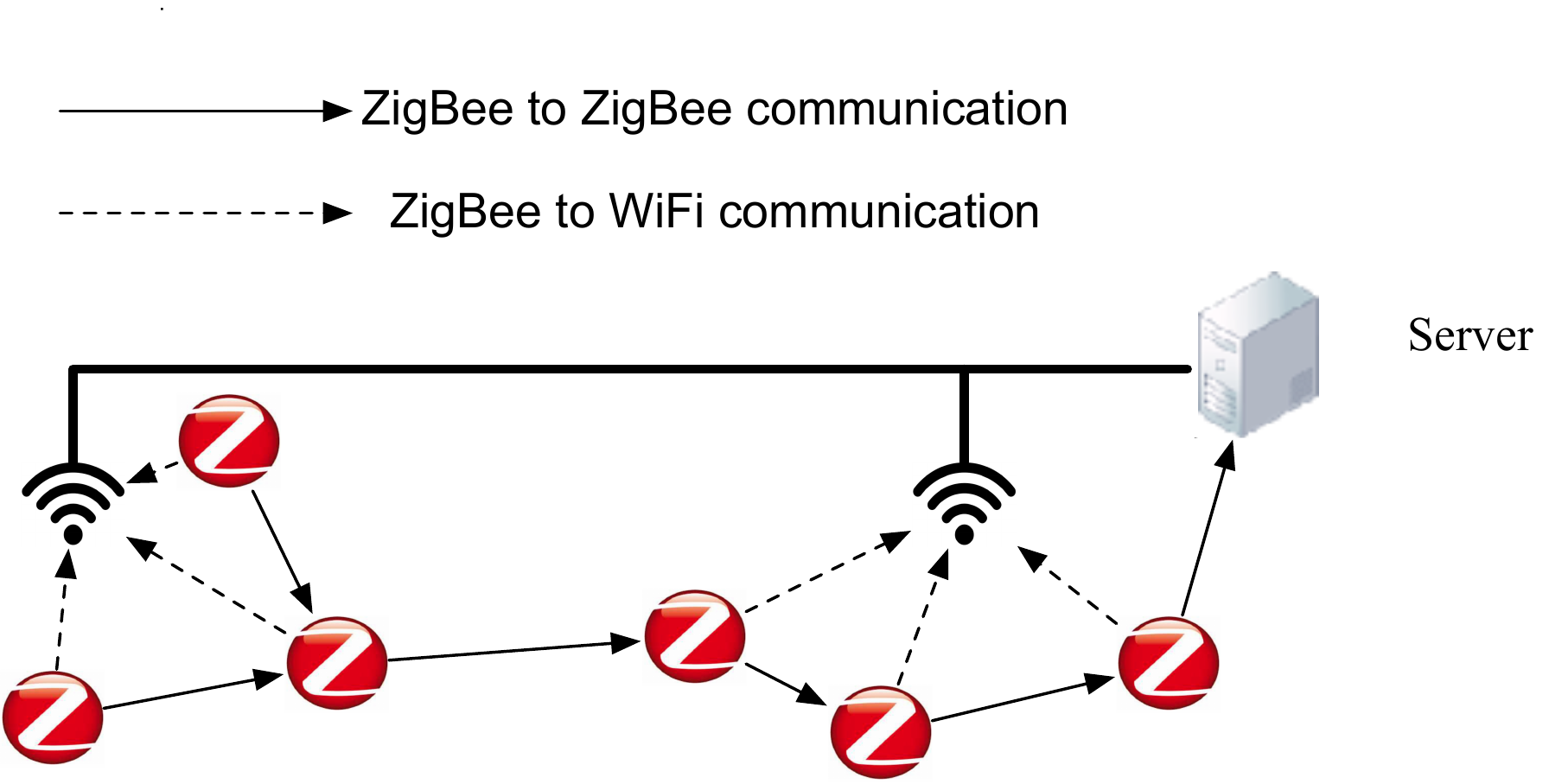}
\caption{The IoT networks model in ECT}
\label{fig:ect}
\end{figure}

\textbf{NetCTC} \cite{netctc} proposes a networking support design for physical layer CTC between WiFi and ZigBee to establish feedbacks.
\textbf{ECT} \cite{ect} proposes a cross-technology network layer design between WiFi and ZigBee. It is a data forwarding method based on B2W2 \cite{B2W2} and EMF \cite{emf}. Fig. \ref{fig:ect} shows ECT's cross-technology network model. When a ZigBee node transmits raw data to other ZigBee nodes, important data is concurrently transmitted to WiFi. Since WiFi APs have direct connections to the server, the priority map generated on the server can be obtained by the WiFi APs. Then, the WiFi APs will broadcast the priority map to ZigBee nodes so that they can forward the data based on their senders' priorities. In this way, ECT dynamically adjusts the priorities of ZigBee nodes and therefore reduces the delivery delay.
\textbf{CRF} \cite{crf} presents a coexistent routing and flooding protocol, which concurrently conducts routing within the WiFi network and flooding among ZigBee nodes using a single stream of WiFi packets. 


\begin{figure}[!tb]
\centering
\includegraphics[width=3.3in]{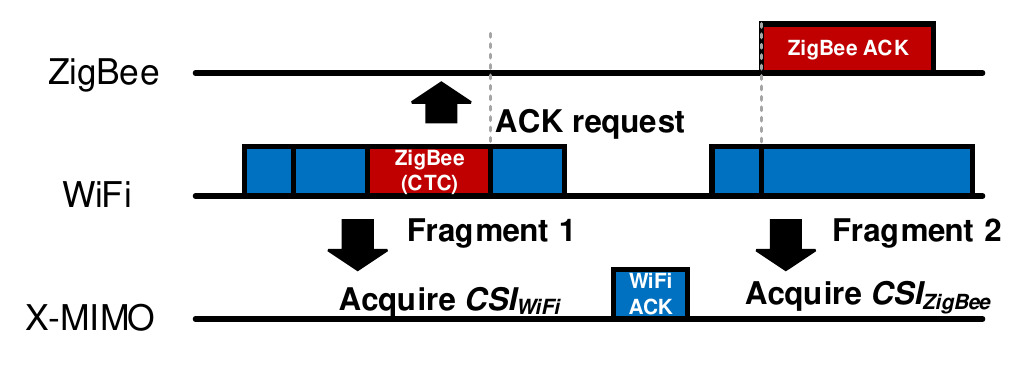}
\caption{Timing control via fragmentation of X-MIMO}
\label{fig:xmimo}
\end{figure}

\textbf{C-LQI} \cite{clqi} proposes a new link metric and a joint link model that considers both the emulation error and the channel distortion. Based on the link model, a lightweight approach is proposed to estimate the quality of the CTC link. C-LQI is defined as the expected probability for a symbol to be correctly decoded by the receiver of a CTC link. 
\textbf{X-MIMO} \cite{bib:X_MIMO} also performs link quality estimation between WiFi and Zigbee to achieve Multi-User Multiple-Input Multiple-Output (MU-MIMO) CTC. It turns the WiFi AP into an MU-MIMO transmitter, delivering different packets to multiple ZigBee devices in parallel. Cross-technology channel estimation is indispensable as the ZigBee channel information must be collected at X-MIMO to support implicit MU-MIMO. As shown in Fig. \ref{fig:xmimo} X-MIMO utilizes a WiFi device associated with the X-MIMO WiFi AP and the WiFi fragmentation function to precisely control the timings of WiFi and ZigBee packets to make them overlap in time. Then the ZigBee channel information can be recovered from this CSI measurement.
With the ZigBee channel information and the ZigBee packets to be sent, X-MIMO precodes these different ZigBee packets into multiple streams. Then the different packets can be decoded by these ZigBee devices simultaneously. The precoded streams are converted into a WiFi packet by multi-stream CTC. This WiFi packet is transmitted by X-MIMO through multiple antennas and is decoded into the different ZigBee packets by the ZigBee devices simultaneously.


\section{Future work}
\label{sec:future}
In order to achieve a more ubiquitous cross-network, cross-frequency, and cross-media communication system, there have also been a lot of cutting-edge works to open a new direction for the development of CTC.

\subsection{Cross-network Systems}
\label{sec:backscatter_crossNetwork}
\begin{figure}[!tb]
\centering
\includegraphics[width=3.1in]{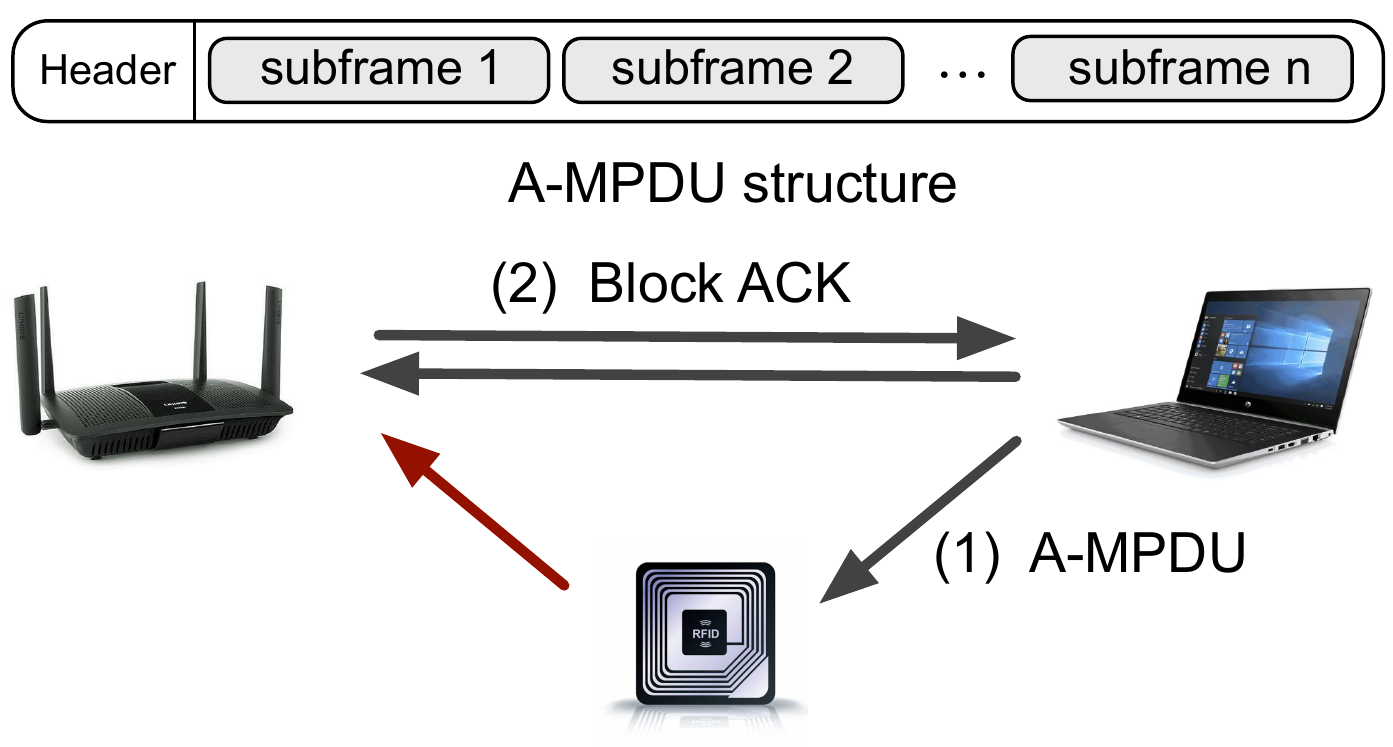}
\caption{WiTAG overview}
\label{fig:WiTag_scenario}
\end{figure}

CTC enables interpreting signals from other technologies with dedicated hardware as a helper in a backscatter communication system~\cite{bib:aloba, bib:aloba_jrnl, bib:Saiyan}. Although a basic backscatter communication system consists of dedicated hardware, it can still be a promising alternative to connect different technologies due to its low power consumption and simplicity. By leveraging the receiver-transparent CTC technique, \textbf{WiTAG} \cite{WiTAG}, shown in Fig. \ref{fig:WiTag_scenario}, designs a MAC-layer backscatter tag that can be read using WiFi devices. It reduces the complexity and cost of deploying backscatter systems by using existing WiFi infrastructures instead of specialized readers. \textbf{Gatescatter} \cite{GateScatter} is another backscatter-based CTC connecting commodity IoT to WiFi. The Gatescatter tag can reshape ZigBee packets with an arbitrary payload into a legitimate 802.11b WiFi packet, such that the payload can be decoded at the WiFi receiver. 

\subsection{Cross-frequency BackCom-based Systems}
\label{sec:backscatter_crossFrequency}

\begin{figure}[!tb]
\centering
\includegraphics[width=3.1in]{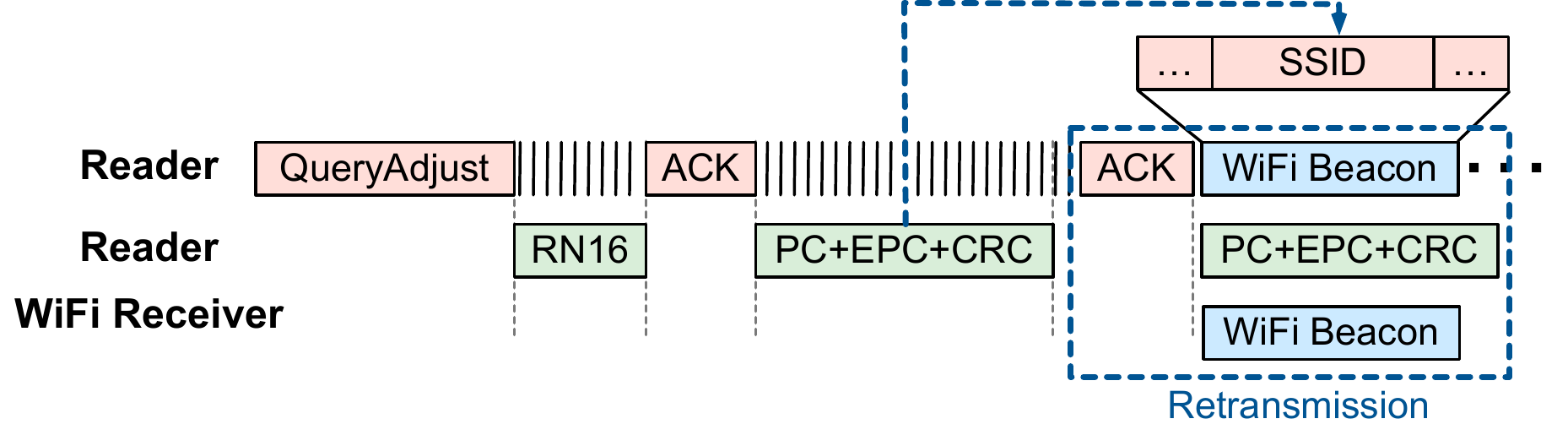}
\caption{Reflection of WiFi beacon in TiFi}
\label{fig:TiFi}
\end{figure}

The above CTC works require the frequency overlapping between the transmitter and the receiver. Nowadays, some CTC works achieve cross-frequency communication between different devices.
\textbf{Interscatter} \cite{interscatter} is the first cross-frequency communication work that achieves CTC from Bluetooth to WiFi in different channels by using backscatter technology. \textbf{TiFi} \cite{TiFi}is another cross-frequency communication work. It achieves CTC from the RFID tag (from 840$MHz$ to 920$MHz$) to the WiFi receiver (2.4$GHz$). TiFi proposes \textit{harmonic backscattering} to sew the frequency gap between RFID and WiFi \cite{Harmonic}. Due to the rectenna's nonlinearity effect, RFID tags reflect the reader's wave at the fundamental frequency (e.g., first at 820$MHz$) and at the harmonics (e.g., second at 1.64$GHz$, third at 2.46$GHz$). In this way, the frequency of the harmonic reflected signal of the RFID tag overlaps with WiFi. The illustration of TiFi process is shown in Fig \ref{fig:TiFi}. Leveraging the RFID mode of retransmission request, the reader can obtain the message of the RFID tag after one transmission. Later, the reader remodulates the message of the RFID tag with the modulation of WiFi. Therefore, the WiFi receiver can receive packets from the RFID tag.

\subsection{Cross-media Systems}
\label{sec:backscatter_crossMedia}

\begin{figure}[!tb]
\centering
\includegraphics[width=2.5in]{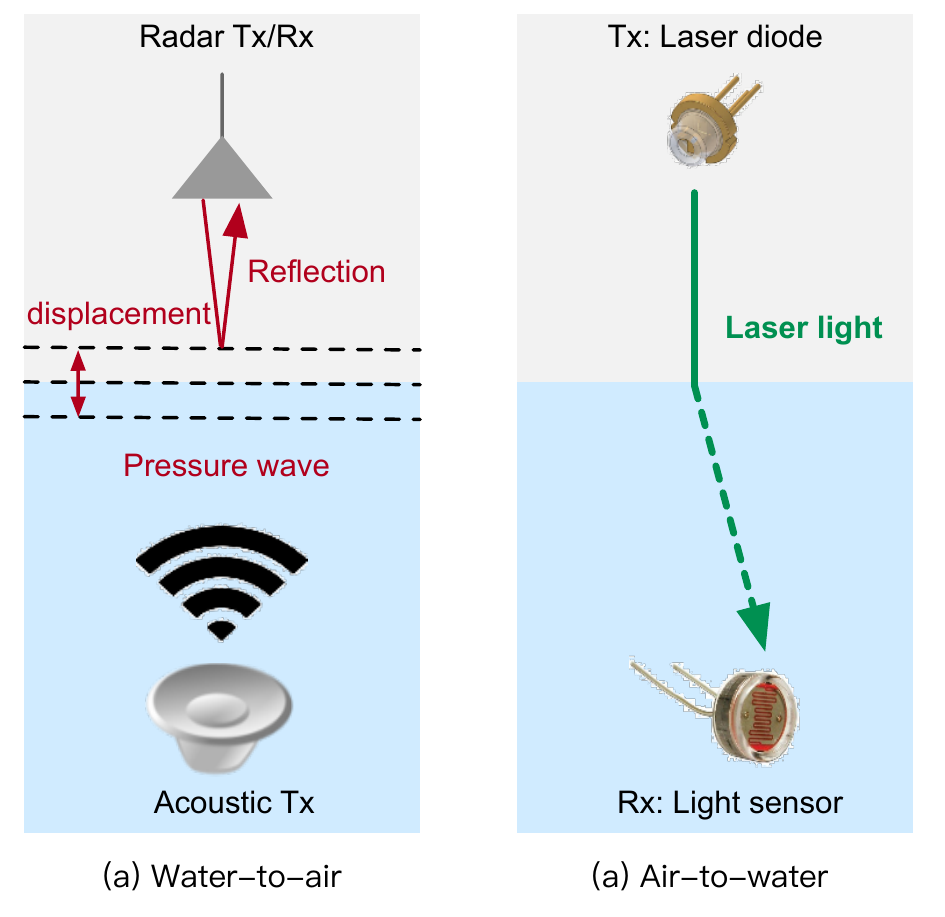}
\caption{Enabling communication across the water-air interface}
\label{fig:cross-media}
\end{figure}

Existing communication technologies cannot enable communication across medium boundaries, such as across the water-air interface. This is because most wireless signals directly reflect at the media boundary. Besides using RF signals, many advanced works explore acoustic and light signals to achieve cross-media communications.
\textbf{TARF} \cite{TARF} enables submerged underwater sensors to directly communicate with an airborne node, as shown in Fig. \ref{fig:cross-media}(a). The design of TARF relies on the fundamental physical properties of acoustic waves. The underwater sensor equipped with a sound transducer transmits the acoustic waves as the pressure waves. When the pressure waves hit the water surface, there will be perturbation or displacement caused by the mechanical nature. In order to pick up the displacement of the surface caused by the acoustic waves, TARF’s airborne sensor transmits frequency-modulated carrier wave (FMCW) millimeter-wave to measure the phase of reflected signals.  \textbf{AmphiLight} \cite{AmphiLight} uses laser light to achieve communication across the air-water interface.

\section{Conclusion}
\label{sec:conclusion}
This paper gives a survey of cross-technology communication in the era of IoT. 
We first point out that the heterogeneity of IoT devices is the critical challenge to achieving communication among heterogeneous wireless devices. 
Such heterogeneity includes the incompatibility of technical standards and the asymmetry of connection capabilities. 
After showing the application potential of CTC, we survey the existing CTC works and classify them into two main categories, packet-level CTC and physical-level CTC.
To have a deep understanding of these works, we analyze those CTC techniques in terms of throughput, reliability, hardware modification, and concurrency. 
We also investigate the works for the construction of upper layers in the CTC network and explore three future trends of CTC: cross-network, cross-frequency, and cross-media. 
Enabling genuinely ubiquitous network connectivity, ubiquitous sensing, and pervasive computing, the CTC technique is deemed an essential component in the infrastructure of future IoT.


\section*{Acknowledgment}
\label{sec:acknowledgement}
This work is supported in part by Joint Funds of the National Natural Science Foundation of China under grant No. U21B2007,
National Science Fund of China under grant No. 61772306 and No, 62072050, China Postdoctoral Science Foundation under grant No. 2021M701888, and the R\&D Project of Key Core Technology and Generic Technology in Shanxi Province (2020XXX007).

%

%
\balance
\bibliographystyle{unsrt}
\bibliography{ref}

\end{document}